\documentclass[conference]{IEEEtran}
\usepackage{graphicx}
\usepackage{balance}  % for  \balance command ON LAST PAGE  (only there!)

 \usepackage{multirow}
\usepackage{scrextend} % for footref
\usepackage{amssymb,amsmath}
\usepackage[figuresright]{rotating}

\usepackage[applemac]{inputenc}

\usepackage{tikz}
\usepackage{pgfplots}

\usepackage{graphicx}% Include figure files
\usepackage{dcolumn}% Align table columns on decimal point
\usepackage{bm}% bold math

\usepackage{amscd,amsthm}
\usepackage{ifthen}
\usepackage{booktabs}

\usepackage[group-separator={,}]{siunitx}

\usepackage{mdframed}
\newcommand{\ignore}[1]{}

\usepackage{tikz}
\usepackage{pgfplots}

\usepgfplotslibrary{groupplots} 
\usetikzlibrary{pgfplots.groupplots}

\usepackage{graphicx}% Include figure files

\usepackage{graphicx}% Include figure files
\usepackage{dcolumn}% Align table columns on decimal point
\usepackage{bm}% bold math
\usepackage{ifthen}
\usepackage{booktabs}

\usepackage{bm}

\usepackage{hyperref}
\hypersetup{
    bookmarks=true,         % show bookmarks bar?
    unicode=false,          % non-Latin characters in Acrobat’s bookmarks
    pdftoolbar=true,        % show Acrobat’s toolbar?
    pdfmenubar=true,        % show Acrobat’s menu?
    pdffitwindow=false,     % window fit to page when opened
    pdfstartview={FitH},    % fits the width of the page to the window
    pdfauthor={Reinhard Heckel},     % author
    pdfsubject={Subject},   % subject of the document
    pdfcreator={Reinhard Heckel},   % creator of the document
    pdfproducer={Producer}, %!TEX encoding = UTF-8 Unicode: producer of the document
    pdfnewwindow=true,      % links in new window
    colorlinks=true,       % false: boxed links; true: colored links
    linkcolor=red,          % color of internal links
    citecolor=blue,        % color of links to bibliography
    filecolor=magenta,      % color of file links
    urlcolor=cyan           % color of external links
}

\usepackage{rays_defs_13}

\definecolor{myred}{RGB}{255,133,133}
\definecolor{myblue}{RGB}{100,200,255}
\definecolor{darkbrown}{RGB}{204,170,153}
\definecolor{lightbrown}{RGB}{221,204,187}
\definecolor{gray}{RGB}{221,221,221}
\definecolor{textgray}{RGB}{70,70,70}

\definecolor{orange}{RGB}{255,153,0}

\definecolor{darkblue}{RGB}{74,120,156}
\definecolor{lightblue}{RGB}{133,193,245}
\definecolor{beige}{RGB}{244,186,112}
\definecolor{mygreen}{RGB}{127,202,159}
\definecolor{myred1}{RGB}{233,109,99}
\definecolor{black}{RGB}{20,20,20}

\pgfplotscreateplotcyclelist{colordashed}{%
myred, very thick\\%
myblue, very thick\\%
beige,very thick, dotted\\%
myred, very thick, dashdotted\\%
black, thin\\%
mygreen, thick, dashed\\%
}
 \pgfplotsset{cycle list name=colordashed}

\newcommand\ind[1]{1_{\{#1 \} }}

\begin{document}

\title{Scalable and interpretable product recommendations \\
via overlapping co-clustering}

\author{\IEEEauthorblockN{Reinhard Heckel}
\IEEEauthorblockA{ University of California \\ Berkeley}
\and
\IEEEauthorblockN{Michail Vlachos}
\IEEEauthorblockA{IBM Research - Zurich}
\and
\IEEEauthorblockN{Thomas Parnell}
\IEEEauthorblockA{IBM Research - Zurich}
\and 
\IEEEauthorblockN{Celestine Duenner}
\IEEEauthorblockA{IBM Research - Zurich}
}

%\author{Reinhard Heckel and Michail Vlachos
%\IEEEcompsocitemizethanks{\IEEEcompsocthanksitem R.  Heckel and M. Vlachos are with the Cognitive Systems group at IBM Research - Zurich.\protect
%}}
% \numberofauthors{2} 
% \author{
% \alignauthor
% Reinhard Heckel\titlenote{Work conducted while at IBM Research, Zurich}\\
%       \affaddr{University of California, Berkeley}\\
%       \affaddr{Dept. of Electrical Eng.}\\
%       \email{heckel@eecs.berkeley.edu}
% % 2nd. author
% \alignauthor
% Michail Vlachos\\
%       \affaddr{IBM Research, Zurich}
% }

% \additionalauthors{Additional authors: John Smith (The Th{\o}rv\"{a}ld Group, {\texttt{jsmith@affiliation.org}}), Julius P.~Kumquat
% (The \raggedright{Kumquat} Consortium, {\small \texttt{jpkumquat@consortium.net}}), and Ahmet Sacan (Drexel University, {\small \texttt{ahmetdevel@gmail.com}})}
% \date{30 July 1999}
% Just remember to make sure that the TOTAL number of authors
% is the number that will appear on the first page PLUS the
% number that will appear in the \additionalauthors section.

\maketitle

%\IEEEtitleabstractindextext{%
\begin{abstract}
We consider the problem of generating interpretable recommendations 
%in a business setting 
% \RH{ we don't have to restrict ourself to the B2B case, we can say that this is our main interrest, but the problem arrises in other applications of recommender systems as well..}
by identifying overlapping co-clusters of clients and products, based only on positive or implicit feedback. Our approach is applicable on very large datasets because it exhibits 
%\RH{we should say almost linear complexity, as the number of iterations of coordinate descent to converge is not independent of the problem size, the dependence is only very mild}
almost linear complexity in the input examples and the number of co-clusters.
We show, both on real industrial data and on publicly available datasets, 
that the recommendation accuracy of our algorithm is competitive to that of state-of-art matrix factorization techniques. In addition, our technique has the advantage of offering recommendations that are textually and visually interpretable. 
Finally, we examine how to implement our technique efficiently on Graphical Processing Units (GPUs).
% and demonstrate a speedup of more than 50 times.
% \RH{I don't think we should talk here about the speedup, because the speedup is relative to our own implementation, and not relative to other peoples or well accepted work. We should just point out that we have a very fast implementation of the algorithm.}
\end{abstract}
%}

%\RH{Let us try to use the following consistently: positive ratings ($r_{ui}=1$), unknown ratings ($r_{ui}=0$) and negative ratings (this would be $r_{ui}=-1$, but since we never actually see negative ratings that does not matter). When we speak of the data that is given to us, we can also speak of positive examples. }

\section{Introduction}
Research on recommender systems is slowly transitioning from its historical focus on prediction accuracy towards a more balanced approach between accuracy and interpretability. Interpretability adds trust, confidence \cite{taxonomyExplanations2011}, and persuasiveness \cite{pursuasive2013} to the recommendations, and in certain scenarios helps highlight tradeoffs and compromises \cite{opinionatedExplanations} and even incorporates aspects of social endorsement \cite{WhoLikesIt2014}. Therefore, interpretable recommendations are of interest in many application areas. 
Here, our goal is to provide interpretable client-product recommendations in a business-to-business (B2B) scenario. In this scenario, the recipient of the recommendation is a salesperson responsible for the client, and not the client itself. A salesperson is responsible for tens or hundreds of clients and to decide whether to pursue a sales opportunity (i.e., recommendation), he or she relies on evaluating the reasoning provided by a generated recommendation.

In a business-to-business, and many other recommender systems, often only positive ratings are present: the products that the clients have already purchased. Negative ratings are unavailable, because absence of a purchase does not necessarily reflect a lack of interest in the item. The problem of generating recommendations based on positive ratings of users only is known as One-Class Collaborative Filtering (OCCF). 
The lack of negative examples makes the problem challenging, as one has to learn the customer's preferences from what they \textit{like}, without information on what they \textit{dislike}. 
In fact, it can be shown that in an online setup, learning only from what users \textit{like} requires more examples for making ``good'' recommendations~\cite{heckel_sample_2017} than learning from both what users like and dislike. 
This problem also occurs in other important collaborative filtering settings, in particular when only implicit ratings are available. Examples of implicit ratings are the browsing history of a visitor at an e-commerce website or views of videos on a video-sharing platform.

State-of-the-art approches to generate recommendations from positive ratings only are often based on standard matrix factorization. However, they offer low interpretability because ``latent factors obtained with mathematical methods applied to the user-item matrix can be hardly interpreted by humans'' \cite{DBLP:conf/dexaw/RossettiSZ13}. Similar observations have been made in several works \cite{nonInterpretable1, nonInterpretable2, Zhang:2014:EFM:2600428.2609579}.

Our approach to providing interpretable recommendations from positive examples is based on the detection of \textit{co-clusters} between users (clients) and items (products)\footnote{\scriptsize In what follows, we use `item' for `product' and `user' for `client'/`company', to conform with the standard literature on recommender systems}. 
Co-clusters are groups of both users and items with similar patterns. Users can have several interests, and items might satisfy several needs, so users and items may belong to several co-clusters. So, co-clusters may be overlapping. Most importantly, discovery of the overlapping user-item co-clusters offers an interpretable model: Identification of sets of users that are interested in or have bought a set of items not only allows us to infer latent underlying patterns but can also lead to better, textually and visually interpretable recommendations. 

\begin{figure}
\includegraphics[width = 0.99\linewidth]{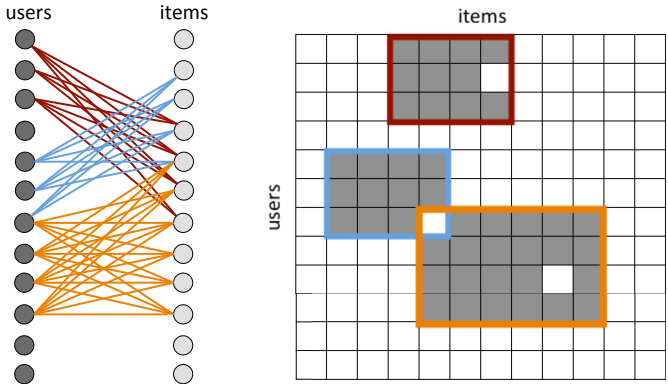}
\caption{\label{fig:ovcomm}Example of overlapping user-item co-clusters identified by the OCuLaR algorithm. Dark squares correspond to positive examples, and the white squares within the clusters correspond to recommendations of the OCuLaR algorithm. }
\end{figure}

Co-clustering principles have been considered before in the context of recommender systems \cite{DBLP:conf/cikm/VlachosFMKV14, khoshneshin2010incremental}, but past work has focused on \emph{non-overlapping} co-clusters, that is, every item/user pair belongs to only one co-cluster. 
%In fact, in the traditional co-clustering setting,  co-clusters are also non-overlapping \cite{dhillon_co-clustering_2001}. 
In contrast, in this work we use  \textit{overlapping co-clusters} to generate interpretable recommendations.
Allowing co-clusters to overlap is not only crucial for the recommendation performance, but also enables the discovery of interesting and useful buying patterns.
In Figure \ref{fig:ovcomm}, we provide an example of overlapping co-clusters. 
A dark square describes a product bought by the user in the past. 
One can visually identify three potential recommendations indicated by white squares inside the co-clusters.
The approach we propose identifies overlapping user-item co-clusters and generates recommendations based on a generative model. 
 The algorithm is scalable,  on par with state-of-art OCCF methodologies in terms of recommendation accuracy, and at the same time provides interpretable recommendations. 
 The algorithm's computational complexity is very low, specifically it is essentially linear in the problem size. We have also developed highly efficient GPU implementations of the algorithm.

%%%%%%%%%%%%%%%%%%%%%%%%%%%%%%%%%%%%%%%%
%%%%%%%%%%%%%%%%%%%%%%%%%%%%%%%%%%%%%%%%

\medskip
\noindent \textbf{Outline:} 
The paper first reviews related work in Section  \ref{sec:related}. Next, we present
our Overlapping co-CLuster Recommendation algorithm, henceforth referred to as ``OCuLaR". 
We start by introducing a generative model, based on which we design a computationally efficient algorithm to produce recommendations and to identify the co-clusters. 
In Section \ref{sec:mtxfact}, we explain the relation to traditional matrix factorization approaches and in Section \ref{sec:gpu} we discuss how to port our approach onto GPUs.
Section \ref{sec:exper} compares OCuLaR to other state-of-the-art recommendation algorithms on a real-world client-product dataset from our institution and publicly available datasets. Our algorithm performs as well as or better than state-of-the-art matrix factorization techniques, with the added benefit of interpretability. 
We conclude with a deployment of our algorithm in a B2B recommender system at our organization.

\section{Related Work}
\label{sec:related}

\medskip\noindent
\textbf{Collaborative Filtering (CF):}
Early approaches to recommender systems performed either user-based or item-based collaborative filtering. User-based techniques infer preferences of a given user based on the preferences of similar or like-minded users by, e.g., by recommending products that the nearest neighbors of a user have bought in the past. 
Similarly, item-based techniques exploit item-to-item similarities to generate recommendations. 
Item- and user-based techniques yield a reasoning of the sort ``similar users have also bought'', but are often outperformed by latent factor models, such as matrix factorization approaches \cite{DBLP:journals/computer/KorenBV09}. Matrix factorization techniques in their traditional form predict ratings or preferences well, but the latent features make it difficult to explain a recommendation \cite{nonInterpretable1, nonInterpretable2,DBLP:conf/dexaw/RossettiSZ13}.
Recently, \cite{Zhang:2014:EFM:2600428.2609579} investigated a method for explainable factorization by extracting explicit factors (sentiment, keywords, etc) from user reviews; however such an approach is applicable only in the presence of additional textual information for each training example. Such data, however, are rarely available in a B2B setting. Generic methodologies for communicating explanations to the user have been studied in \cite{Tintarev:2007:SER:1547550.1547664, taxonomyExplanations2011, Gedikli2014}. 

%Explanations in RSs \cite{Tintarev:2007:SER:1547550.1547664} \cite{Gedikli2014} \cite{Tintarev2012}

Conceptually, the algorithm proposed in this work can be regarded as a combination of item-based and user-based approaches. 
Specifically, we discover similarities that span across both the user and the item space by identifying user-item co-clusters. 
By identifying user-item co-clusters via matrix factorization techniques, we obtain an estimate of the probability that an unknown example is positive, which automatically results in ranked recommendations. 

\textit{OCCF models} are used to predict preferences from implicit feedback. This is a typical recommendation scenario when browsing, buying, or viewing history is available, or in general in setups where the user does not provide any explicit negative feedback. Recommendation techniques based on positive ratings try to learn by considering the implicit feedback either as absolute preferences \cite{pan_one-class_2008,Koren2008}, or as relative preferences \cite{Rendle2009, GBPR}. 
%In this case, a user $u$ is assumed to prefer item $i$ over item $j$ if the user-item pair $(u,i)$ is among the positive examples (e.g., $i$ has been purchased by $u$), and $(u,j)$ is not among the positive examples (e.g., $j$ has not been purchased by $u$).
We compare with techniques from both categories in Section \ref{sec:exper}, and state the formal relation of our approach to matrix factorization techniques in Section \ref{sec:mtxfact}.

\medskip\noindent
\textbf{Co-clustering:}
The majority of the literature on co-clustering considers the detection of non-overlapping co-clusters \cite{dhillon2003information, papadimitriou2008disco}. 
Notable exceptions are the approach in \cite{Banerjee:2005:MOC:1081870.1081932}, which introduces a generative model for overlapping co-clusters along with a generic alternating minimization algorithm for fitting the corresponding model to given data, and the approach in 
\cite{DBLP:conf/icdm/ShafieiM06}, which considers the problem of simultaneously clustering documents and terms. That latter approach is based on a generative model and estimation of the corresponding model parameters \cite{DBLP:conf/icdm/ShafieiM06}. In both \cite{Banerjee:2005:MOC:1081870.1081932,DBLP:conf/icdm/ShafieiM06}, the focus is on discovering co-clusters, whereas here we are interested in explaining the recommendations produced by our algorithm with co-clusters. 
Although co-clustering approaches have been used before in the collaborative filtering setting, the majority of those papers \cite{george2005scalable,khoshneshin2010incremental,DBLP:conf/cikm/VlachosFMKV14} is restricted to non-overlapping co-clusters.
An exception is \cite{xu2012exploration}, which explores the multiclass co-clustering problem in the context of CF.

\medskip\noindent
\textbf{Community detection:}
Related to co-clustering is community detection. To see this, observe that the positive examples are the edges in a bipartite graph of users and items (see Figure \ref{fig:ovcomm}). 
Then, the co-clusters of users and items correspond to user-item communities in the bipartite graph. 
One of the best-known community detection algorithms is based on the notion of modularity \cite{Newman2006}. 
Specifically, the modularity algorithm by Girvan \& Newman \cite{Girvan2002} is one of the most widely-used community detection algorithms and is used in many software packages (gephy, Mathematica, etc). 
It has the advantage that it can automatically discover the number of communities; however it does not support discovery of overlapping communities. 
Other work on detecting non-overlapping communities includes  \cite{karypis1998multilevelk,dhillon2007weighted}.

\begin{figure}[!ht]
\centering
\begin{tabular}{l r}
\includegraphics[width=.46\linewidth]{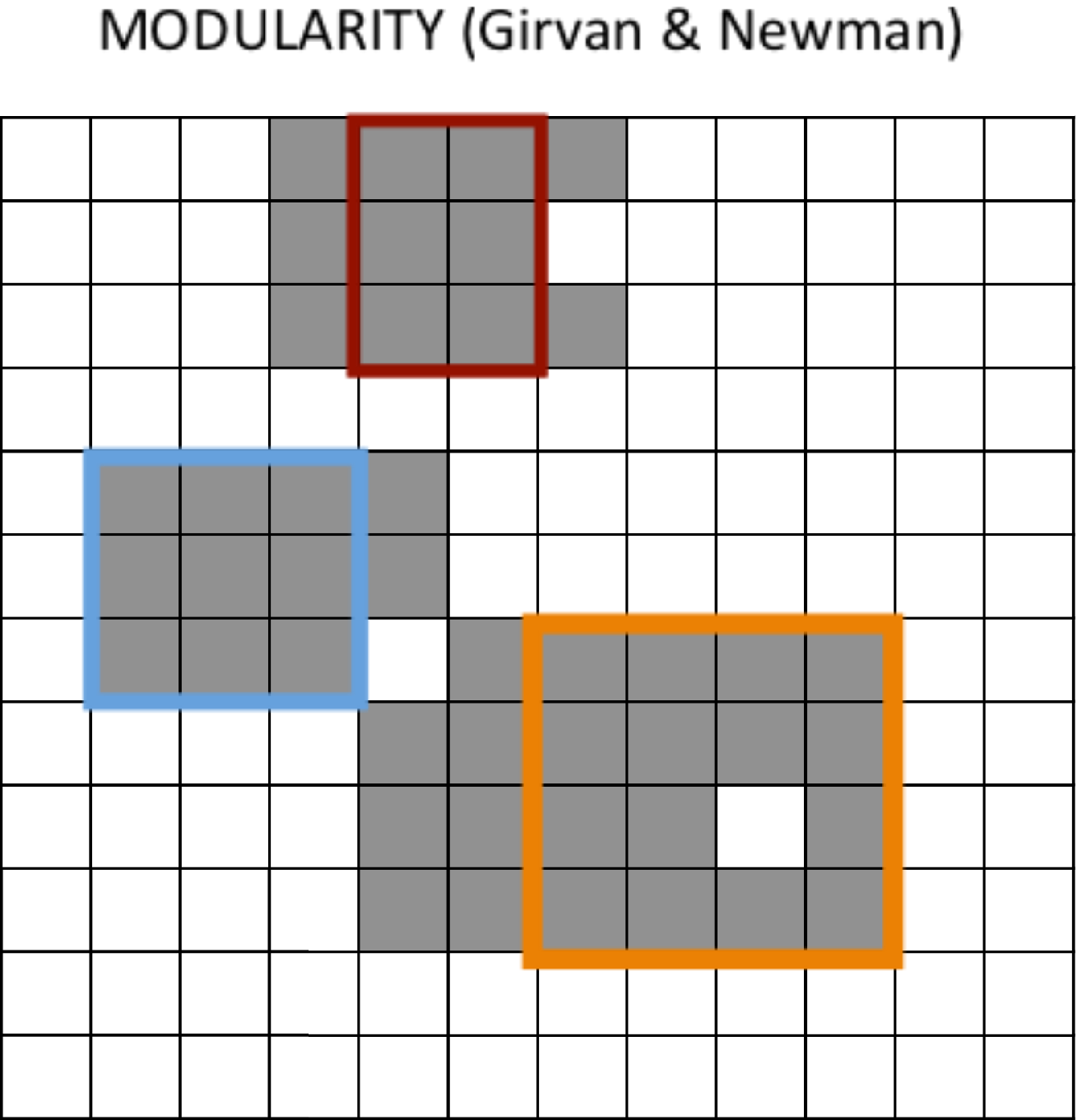} &  
\includegraphics[width=.46\linewidth]{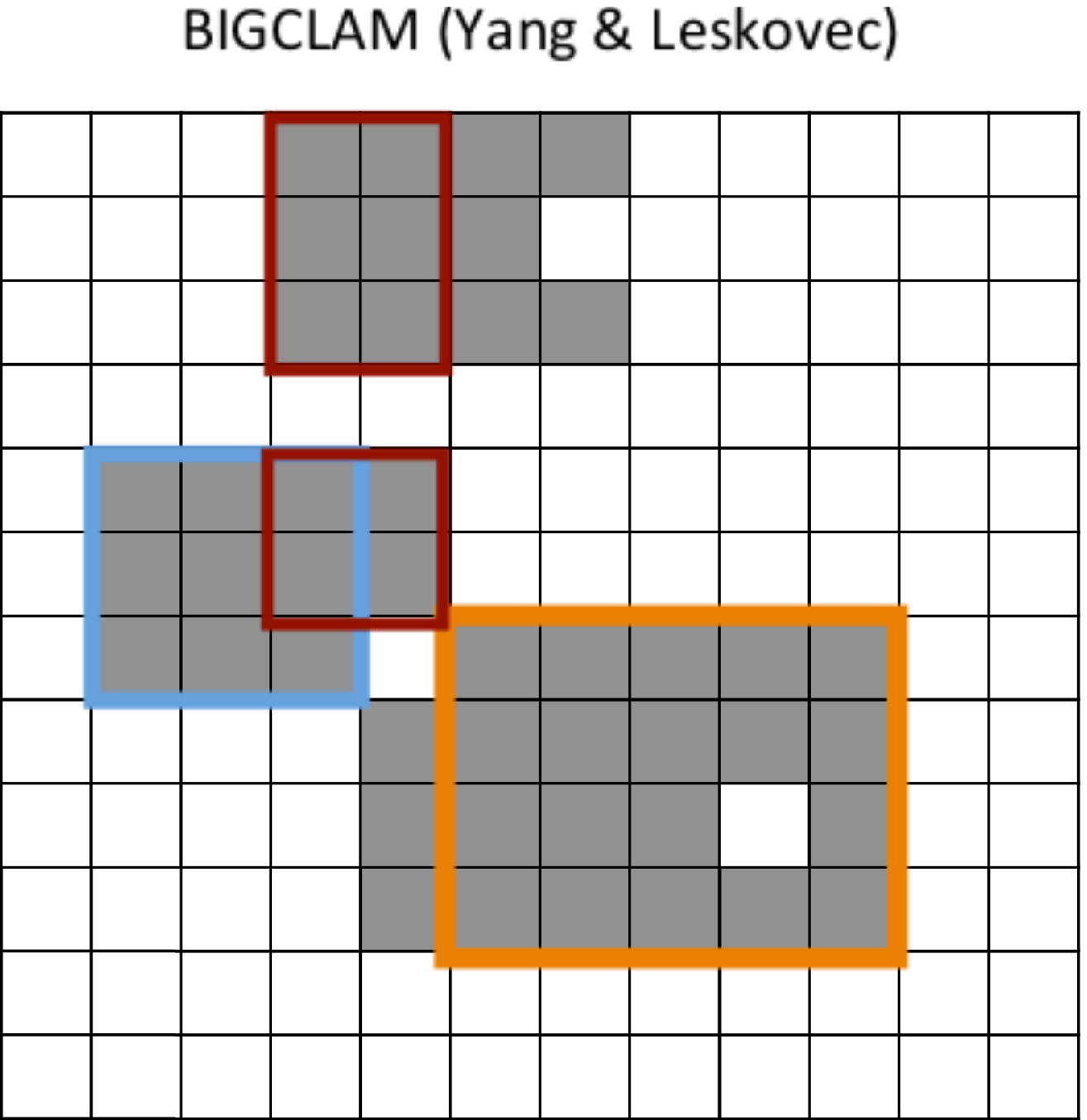}  
\end{tabular}
%\vspace{-\baselineskip}
\caption{Output of non-overlapping (Modularity) and overlapping (BIGCLAM) community detection algorithms.
We see that both fail to recover the correct community structure.}
\label{fig:intro_graph_other}
\end{figure}

 Recently, interest in identification of overlapping communities has grown  \cite{palla2005uncovering,airoldi2009mixed,leskovec2012learning,bigclam, DBLP:journals/corr/Peixoto14}.
% \RH{ The sentence below is important to give credit. }
 Most related to our work is the BIGCLAM algorithm proposed by Yang and Leskovec \cite{bigclam}; the main differences are that in our approach we consider a particular bi-partite graph and use regularization, which turns out to be crucial for recommendation performance.  
While such approaches share similar concepts, off-the-shelf community detection methodologies are not directly applicable to the one class  collaborative filtering problem, as they yield an assignment of users/items to communities, but not a ranked list of recommendations. In fact, they may even fail to reveal the correct community structure. As an illustrative example, in Figure \ref{fig:intro_graph_other} we provide the output of both Modularity and BIGCLAM for the introductory example. Both fail to reveal the correct co-clustering structure, and by recovering incorrect `community' boundaries they would have identified only one (1) of the three (3) candidate recommendations.

%%%%%%%%%%%%%%%%%%%%%%%%%%%%%%%%%%%%%%%%%%%%%%%%%%%%%%
%%%%%%%%%%%%%%%%%%%%%%%%%%%%%%%%%%%%%%%%%%%%%%%%%%%%%%

\section{B2C vs B2C recommendations}

Our main application of interest, albeit by far not the only one, in the problem of generating interpretable recommendation from one-class examples are B2B recommender systems. 
A B2B recommender system---in contrast to a Business-to-Consumer (B2C) recommender system (referring to recommender systems employed at sites such as Amazon.com or the Genius recommendations of Apple within their iTunes platform)---is typically not open to the public, but is deployed within an enterprise. Because only a handful of literature is available on B2B recommender systems \cite{DBLP:conf/icwe/OpreaHZEL13,DBLP:conf/iat/ZieglerOHEL13, B2B_ecommerce},
 we briefly point out some differences between a B2C and B2B deployment.
This discussion will serve the purpose of highlighting the need for higher interpretability in a B2B recommendation environment. %The discussion will also help highlight the different assumptions in data availability for both systems. 
%\RH{Commented this out, as the problem really arises in both B2B as well as B2C environments..}

\begin{enumerate}

\item The \textbf{system users} in a B2C system are the consumers, whereas in B2B they are the selling teams or marketing teams of an enterprise. Therefore the recommendations have a different audience.

\item The \textbf{user entities} (``clients") for which the recommendations are created are typically individuals in a B2C setting. In B2B, they are enterprises or companies. This has certain implications on data availability. For example, in B2C we have knowledge about past buying patterns (PBP) or implied interest from items browsed by the users. In B2B, we know the past buying patterns and it is also possible to obtain other type of information about the clients-companies in the form of news events or financial reports published about that company. This eventually can help build a more holistic view of the examined bipartite graph of clients and products.

\item The \textbf{value of the recommendation} in a B2C setting is typically small, ranging from a few dollars (a song or a movie) to a few hundreds of dollars (an electronic gadget). In contrast, in a B2B setting, the recommendation may refer to a large hardware installation or complex and custom software solution and the recommendation may be valued at thousands of dollars, or even millions. It is also interesting to note that in B2C the value of a product/recommendation  is (mostly) fixed, or it is publicly available, whereas in B2B it varies based on the complexity of the installation.

\item Finally, the \textbf{reasoning} of the recommendation is short and limited in a B2C scenario, but it has to be detailed and offer sufficient supportive evidence in B2B. This is partially because of the associated value of the recommendation. If a salesperson is to pursue a multi-thousand dollar sales opportunity, the methodology has to present sufficient reasoning for this action, which is subsequently further evaluated by the responsible seller or marketing team.

\end{enumerate}

%These points are summarized in Table \ref{tab:differences}. 
In this work we will only focus on the last part, the reasoning of the recommendation. Even though components such as estimating the appropriate value of a B2B recommendation are also non-trivial, they will not be further examined in this work.

% \begin{table*}[]
%     \centering
%     \caption{Differences between a B2C and B2B recommender system}
%     \label{tab:differences}
%     \small
%   \centering
%  \resizebox{.98\columnwidth}{!}{%
%     \begin{tabular}{l l l}
%          & \textbf{B2C}  & \textbf{B2B} \\  \hline
%         \textbf{users of system} & consumers & sellers/marketing teams\\
%          \textbf{``clients"}       & consumers & enterprises/companies \\
%          \textbf{what we know about clients}            & past buying patterns (PBP) & PBP, news, financials,... \\
%          \textbf{value of recommendation} & small to medium &  medium to high \\
%          \textbf{reasoning of recommendation} & limited  & detailed \\ 
%     \end{tabular}
%     } %resizebox
% \end{table*}

In the following section we begin to describe our approach for generating interpretable recommendations based on the client-product purchase graph.

%%%%%%%%%%%%%%%%%%%%%%%%%%%%%%%%%%%%%%%%%%%%%%%

%\vspace{-.5\baselineskip}
%\section{Formal problem statement}

\section{OCuLaR algorithm}
\label{sec:ocular}
%In this section we state our Overlapping co-CLuster Recommendation (OCuLaR) algorithm which is based on matrix factorization principles.

Now we present our Overlapping co-CLuster Recommendation algorithm, or for short ``OCuLaR".  We assume that we are given a matrix $\mR$ where the rows correspond, e.g., to users or clients and the columns to items or products. 
If the $(u,i)$th element of $\mR$ takes on the value $r_{ui} =1$ this indicates that user $u$ has purchased item $i$ in the past or, more generally, that user $u$ is interested in item $i$. 
We consider all values $r_{ui}$ that are not positive ($r_{ui} =1$) as unknown ($r_{ui}=0$),  because they indicate that user $u$ might be interested in $i$ or not. 
Our goal is to identify those items a user $u$ is likely to be interested in.
Put differently, we want to find the positives among the unknowns, given only positive examples.

We assume an underlying model whose parameters are factors associated with the users and items. Those factors are learned, such that the fitted model explains well the given positive examples $r_{ui}=1$.
The specific choice of our model 
 %has the advantage that, in order to learn its parameters, the corresponding likelihood takes on a convenient form, 
allows us to design an efficient algorithm to learn the factors. 
%which allows to employ an efficient algorithm. 
Moreover, the factors encode co-cluster membership and affiliation strength, and are interpretable in that sense. 

\subsection{Generative model}
\label{seg:genmod}
We start with the generative model underlying our recommendation approach. 
It formalizes the following intuition: There exist clusters, groups, or communities of users that are interested in a subset of the items.
As users can have several interests, and items might satisfy several needs, each user and item can belong to several co-clusters consisting of users and items. 
However, a co-cluster must contain at least one user and one item, and can therefore not consist of users or items alone.

Suppose there are $K$ co-clusters ($K$ can be determined from the data, e.g., by cross-validation, as discussed later). 
Affiliation of a user $u$ and item $i$ with a co-cluster is modeled by the $K$-dimensional co-cluster affiliation vectors $\vf_u$ and $\vf_i$,  respectively. 
The entries of $\vf_u, \vf_i$ are constrained to be non-negative, and  $[\vf_u]_c = 0$ signifies that user $u$ does not belong to co-cluster $c$. Here, $[\vf]_c$ denotes the $c$-th entry of $\vf$. 
The absolute value of $[\vf_u]_c$ corresponds to the affiliation strength of $u$ with co-cluster $c$; the larger it is, the stronger the affiliation. 

Positive examples are explained by the co-clusters as follows. If user $u$ and item $i$ both lie in co-cluster $c$, then this co-cluster generates a positive example with probability 
\[
1 - e^{-[\vf_u]_c [\vf_i]_c }. 
\]
Assuming that each co-cluster $c=1,...,K$, generates a positive example independently, it follows that 
%If user $u$ did not purchase $i$, they should not lie in any common community, i.e., 
\[
1 - \PR{r_{ui} = 1}  = \prod_c e^{-[\vf_u]_c [\vf_i]_c } = e^{- \innerprod{\vf_u}{\vf_i}},
\]
where $\innerprod{\vf}{\vg} = \sum_c [\vf]_c [\vg]_c$ denotes the inner product in $\reals^K$. Thus 
\begin{align}
\PR{r_{ui} = 1} = 1 - e^{- \innerprod{\vf_u}{\vf_i}}. 
\label{eq:prrui1}
\end{align}
A similar generative model also appears in the community detection literature \cite{bigclam}.
So far the model cannot explain a positive example by means other than co-cluster affiliation. 
Although we could incorporate user ($b_u$), item ($b_i$), and overall bias ($b$) by supposing that the probability of an example being positive is given by 
\[
\PR{r_{ui} = 1} = 1 - e^{- \innerprod{\vf_u}{\vf_i} -b_u - b_i -b },
\]
we found that fitting the corresponding model does not increase the recommendation performance for the datasets considered in Section \ref{sec:exper}, and therefore will not discuss it further.

\newcommand{\setU}{\mc U}
\newcommand{\setI}{\mc I}

\subsection{\label{sec:fittinmp}Fitting the model parameters}

Given a matrix $\mR$, we fit the model parameters by finding the most likely factors  $\vf_u,\vf_i$ to the matrix $\mR$ by maximizing the likelihood  (recall that we assume positive examples to be generated independently across co-clusters and across items and users in co-clusters):
\[
\mc L = 
\prod_{(u,i)\colon r_{ui}=1}
( 1 - e^{ -\innerprod{\vf_u}{ \vf_i} } )
\prod_{(u,i)\colon r_{ui}=0}
e^{ -\innerprod{\vf_u}{ \vf_i} }.
\]
Maximizing the likelihood is equivalent to minimizing the negative log-likelihood:
\begin{small}
\begin{align}
-\log \mc L 
%= - \log \PR{r_{ui} | \mF}
= - \sum_{(u,i)\colon r_{ui} = 1}
\log( 1 - e^{ -\innerprod{\vf_u}{ \vf_i} } )
+ \sum_{(u,i) \colon r_{ui}=0 }  \innerprod{\vf_u}{ \vf_i}.
\label{eq:modlogli}
\end{align}
To prevent overfitting, we add an $\ell_2$ penalty, which results in the following optimization problem:
\begin{align}
\text{minimize }Q \text { subject to } [\vf_{u}]_c, [\vf_{i}]_c  \geq 0, \text{ for all } c,
\label{eq:optprob}
\end{align}
where 
\begin{align}
Q = -\log \mc L + \lambda \sum_i \norm[2]{\vf_i}^2  + \lambda \sum_u \norm[2]{\vf_u}^2 
\end{align}
\end{small}
and $\lambda\geq 0$ is a regularization parameter. As will we discuss in more detail in Section \ref{sec:mtxfact}, this optimization problem can be viewed as a variant of non-negative matrix factorization (NMF), specifically NMF with a certain cost function. 

A common approach to solve an NMF problem is alternating least squares, which iterates between fixing $\vf_u$, and minimizing with respect to $\vf_i$, and fixing $\vf_i$ and minimizing with respect to $\vf_u$, until convergence. 
This strategy is known as cyclic block coordinate descent or the non-linear Gauss-Seidel method. 
Whereas $Q$ is non-convex in  $\vf_i,\vf_u$, $Q$ is convex in  $\vf_i$ (with  $\vf_u$ fixed) and convex in  $\vf_u$ (with $\vf_i$ fixed). 
Therefore, a solution to the subproblems of minimizing $Q$ with fixed $\vf_i$ and minimizing $Q$ with fixed $\vf_u$ can be found, e.g., via gradient descent or Newton's method. 
As this optimization problem is non-convex, one cannot in general guarantee convergence to a global minimum; however convergence to a stationary point can be ensured. 
Specifically, provided that $\lambda > 0$, $Q$ is strongly convex in  $\vf_i$ (with  $\vf_u$ fixed) and in $\vf_u$ (with $\vf_i$ fixed). 
Thus, the subproblems have unique solutions, and therefore, if we solve each subproblem exactly, convergence to a stationary point is ensured by \cite[Prop.~2.7.1]{bertsekas_nonlinear_1999}. 

However, as noted in the context of matrix factorization \cite{hsieh_fast_2011}, solving the subproblems exactly may slow down convergence. 
Specifically, when $\vf_u,\vf_i$ are far from a stationary point, it is intuitive that there is little reason to allocate computational resources to solve the subproblems exactly. 
It is therefore often more efficient to solve the subproblem only approximately in each iteration \cite{bonettini_inexact_2011,hsieh_fast_2011}. 

For the above reasons, we will only approximately solve each subproblem by using a single step of projected gradient descent with backtracking line search, and iteratively update $\vf_i$ and $\vf_u$ by single projected gradient descent steps until convergence. In Section \ref{sec:impdet}, we provide further implementation details. 
Convergence is declared if $Q$ stops decreasing. 
This results in a very efficient algorithm that is essentially linear in the number of positive examples $\{(u,i)\colon r_{ui} = 1\}$, and the number of co-clusters $K$. 
Our simulations have shown that performing only one gradient descent step significantly speeds up the algorithm. 

\medskip\noindent
\textbf{Choice of $K$ and $\lambda$:}
Recall that the number of co-clusters $K$ and the regularization parameter $\lambda$ are the two model hyper-parameters. $K$ and $\lambda$ can be determined from the data via cross-validation. 
Specifically, to determine a suitable pair of ($K$, $\lambda$), we train a model on a subset of the given data for different choices of ($K$, $\lambda$), and select the pair for which the corresponding model performs best on the test set. The model can be tuned for an appropriate metric. In our experiments, we measure the recommendation performance in terms of the recall-at-$M$ items \cite{DBLP:conf/ecir/McSherryN08}, which is a typical performance metric in recommender systems. Because this grid-search approach can be costly if one wishes to perform a fine-grained search of the hyper-parameter space, the GPU implementation of the algorithm, reported in Section \ref{sec:gpu}, can help to dramatically reduce the overall grid-search time.

%%%
\subsection{Generating interpretable recommendations}

Suppose we want to recommend $M$ items to each user. After having fitted the model parameters, we recommend item $i$ to user $u$ if $r_{ui}$ is among the $M$ largest values $\PR{r_{ui'} = 1}$, where $i'$ is over all items that user $u$ did not purchase, i.e., over all $i'$ with $r_{ui'} = 0$. 
%The justification along with the recommendation $i$ to user $u$ is computed as follows. 
The probability $\PR{r_{ui} = 1}$ is large if the user-item pair $(u,i)$ is in one or more  user-item co-clusters. Thus, along with a recommendation, we can output the corresponding user-item co-clusters that cause $\PR{r_{ui'} = 1}$ or, equivalently,  $\innerprod{\vf_u}{\vf_i} = \sum_{c} [\vf_u]_c [\vf_i]_c$ to be large. 
The user-item co-cluster $c$ is determined as the subset of users and items for which $[\vf_u]_c$ and $[\vf_i]_c$, respectively, are large. 

In the B2B setting that we consider it is also important to explicitly mention \textit{who} are the clients that have purchased a similar bundle of products. Contrary to the B2C setting, where, for privacy reasons, one only mentions what \textit{similar} clients purchase, in a B2B setting this is not a concern. The salesperson, who is the recipient of the recommendation, can use this information (explicit names of similar clients) to understand better the types of clients/companies that typically require such a solution. Our approach directly provides this information, because each co-cluster consists of specific clients (users) and products, and does not merely describe an \textit{average} behavior.

%
%\vspace{-\baselineskip}
%%%%%%%%%%%%%%%%%%%%%%%%%%%%%%%%%%%%%%%%%%%%%%%%%%%%%%%%%%%%%%%%%%%%%%%%%%%%
\begin{figure}[ht]
\begin{center}
\includegraphics[width = 0.35\textwidth]{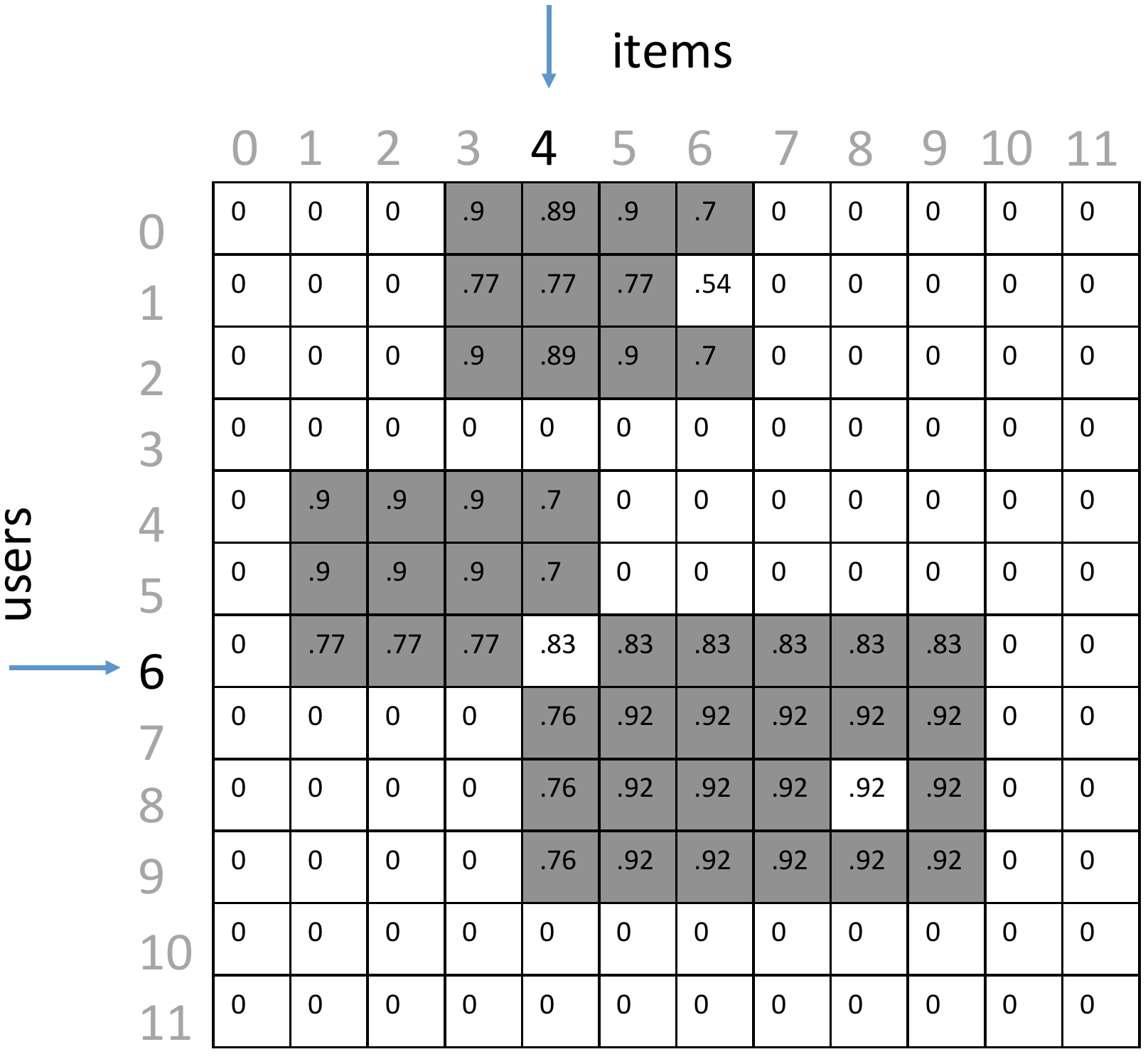}
\caption{\label{fig:probsex} 
Probability estimates $\PR{r_{ui}} = 1 - e^{-\innerprod{\vf_{u}}{\vf_i}}$ of the OCuLaR algorithm applied to the example in Figure \ref{fig:ovcomm}; gray rectangles correspond to positive examples ($r_{ui}=1$), white rectangles to unknown examples ($r_{ui}=0$).
}
\end{center}
\end{figure}
%\vspace{-\baselineskip}
%%%%%%%%%%%%%%%%%%%%%%%%%%%%%%%%%%%%%%%%%%%%%%%%%%%%%%%%%%%%%%%%%%%%%%%%%%%%

%\medskip
%\vspace{-1\baselineskip}
\noindent \textbf{Example:}
Here we provide a cogent example about the interpretability of the recommendations by OCuLaR. We will use the user-item array from Figure \ref{fig:ovcomm}. Let us consider making one recommendation to User 6, where Users 0--11 correspond to the rows, and Items 0--11 to the columns of the matrix. 
The probabilities of the fitted model for each user-item pair are depicted in Figure \ref{fig:probsex}. 
The probability estimate $\PR{r_{ui}} = 1 - e^{-\innerprod{\vf_{u}}{\vf_i}}$,  for $u=6$ is maximized among the unknown examples ($r_{ui}=0$) for Item $i=4$, and is given by $0.83$. 
Therefore, OCuLaR recommends Item $i=4$ to User $u=6$. 
The corresponding factors are $\vf_{i} = [ 1.39 ,  0.73,  0.82]$ and $\vf_{u} = [0 ,  1.05,  1.25]$, which means that Item $i=4$ is in all three co-clusters, while User $u=6$ is in co-clusters 2 and 3 only. 
The probability estimate $1 - e^{-\innerprod{\vf_{u}}{\vf_{i}}}$ for $u=6,i=4$ is large because both User 6 and Item 4 are in co-clusters 2 and 3. 
To justify recommending Item $4$ to User $6$, we can therefore give the following, automatic interpretation to the user of the recommender system: 
%%%%%%%%%%%%%%%%%%%%%%%%%%%%%%%%%%%%%%%

{
%\normalfont\mdseries
%\fontfamily{fi4}\selectfont 
\fontfamily{pag}\selectfont 
\scriptsize
\begin{mdframed}[
  leftmargin=\parindent,
  rightmargin=\parindent,
  skipabove=\topsep,
  skipbelow=\topsep
  ]
Item 4 is recommended to Client 6 with confidence 0.83 because: 
\begin{itemize}
\item Client 6 has purchased Items 1-3. Clients with similar purchase history (e.g., Clients 4-5) also bought Item 4. 
\item Moreover, Client 6 has purchased Items 5-9. Clients with similar purchase history (e.g., Clients 7-9) also bought Item 4.
\end{itemize}
\end{mdframed}
}

Naturally, once the co-clusters have been discovered, additional information (derived from the co-clusters) can be attached in the rationale presented. %In Section \ref{sec:incorpatt}, we also discuss how additional information can be incorporated directly into the OCuLaR model. 
In the experimental Section \ref{sec:exper}, we discuss the recommendation rationale for an industrial deployment of OCuLaR (see Figure \ref{fig:example_s}).

\subsection{Implementation and complexity}
\label{sec:impdet}
Here we examine in more detail the projected gradient descent approach we use to solve the subproblems and the complexity of the overall optimization algorithm. 
% It is important to give credit here:
It is sufficient to discuss minimization of $Q$ with respect to $\vf_i$, as minimization with respect to $\vf_u$ is equivalent.  
The following approach for minimizing $Q$ was also used in similar form in \cite{bigclam, lin_projected_2007}. 
%$-\innerprod{\vq_i}{\vq_u}$ is convex (not strictly convex, of course)
%so 
%\begin{figure}
%\begin{tikzpicture}
%    \begin{axis}[
%        xlabel=$x$,
%        ylabel=$-\log(1-e^{-x})$
%    ]
%    %  invoke external gnuplot as
%    %  calculator:
%    \addplot[mark=none,color=blue] gnuplot[domain=0.01:5,samples=100]{-log(1-exp(-x))};
%    \end{axis}
%\end{tikzpicture}
%
%\begin{tikzpicture}
%    \begin{axis}[
%        xlabel=$x$,
%        ylabel=$1-e^{-x}$
%    ]
%    %  invoke external gnuplot as
%    %  calculator:
%    \addplot gnuplot[domain=0.01:5,samples=100]{1-exp(-x)};
%    \end{axis}
%\end{tikzpicture}
%\end{figure}
%
We start by noting that, because of  
\begin{align*}
Q 
&= 
\sum_{i} \left( 
- \sum_{u \colon r_{ui} = 1}
\log( 1 - e^{- \innerprod{\vf_u}{ \vf_i} }  ) 
+ \sum_{u \colon r_{ui} = 0}  \innerprod{\vf_u}{ \vf_i}  \right) \\
&+ \lambda \sum_{u} \norm[2]{\vf_u}^2 + \lambda \sum_i \norm[2]{\vf_i}^2,
\end{align*}
we can minimize $Q$ for each $\vf_i$ individually. The part of $Q$ depending on  $\vf_i$ is given by
\begin{align}
\begin{small}
Q(\vf_i)
=
- \!\!\!\!\!\sum_{u\colon r_{ui} = 1 }
\log( 1 - e^{- \innerprod{\vf_u}{ \vf_i} }  )
\!+\!  \innerprod{\vf_i }{  \sum_{u \colon r_{ui} = 0 }  \vf_u }  \!+\! \lambda \norm[2]{\vf_i}^2.
\label{eq:coordstep}
\end{small}
\end{align}
As mentioned, we update the parameter $\vf_i$ by performing a projected gradient descent step. 
The projected gradient descent algorithm \cite[Sec.~2.3]{bertsekas_nonlinear_1999} is initialized with a feasible initial factor $\vf_i^{0}$ and updates the current solution $\vf^k_i$ to $\vf^{k+1}_i$ 
according to  
\[
\vf^{k+1}_i = (\vf^{k}_i - \alpha_k   \nabla Q(\vf^k_i) )_+,
\]
where $(\vf)_+$ projects $\vf$ on its positive part, $[(\vf)_+]_c = \max(0, [\vf]_c )$, and the gradient is given by
%The gradient of $Q(\vf_i)$ is: 
\begin{align}
\nabla Q( \vf_i ) 
= 
-\sum_{u\colon r_{ui} = 1}  \vf_u \frac{e^{- \innerprod{\vf_u}{ \vf_i } }}{1 -e^{-  \innerprod{\vf_u}{ \vf_i} }} 
+ \sum_{u\colon r_{ui} = 0} \vf_u
+ 2 \lambda \vf_i.
\label{eq:origgrad}
\end{align}
The step size $\alpha_k$ is selected using a backtracking line search, also referred to as the Armijo rule along the projection arc \cite{bertsekas_nonlinear_1999}. 
Specifically, $\alpha_k = \beta^{t_k}$, where $t_k$ is the smallest positive integer such that 
\[%\begin{align}
Q(\vf^{k+1}_i) - Q(\vf^k_i)
\leq
\sigma \innerprod{\nabla Q(\vf^k_i)}{ \vf^{k+1}_i  - \vf^k_i }
\]
%\label{eq:armijorule}
%\end{align}
where $\sigma,\beta \in (0,1)$ are user-set constants. 
As the computation of both $\nabla Q(\vf_i)$ and $Q(\vf_i)$ requires $\sum_{u \colon r_{ui} = 0} \vf_u$, and typically, the number of items for which $r_{ui}=1$ is small relative to the total number of items, we precompute $
\sum_u \vf_u 
$
before updating all $\vf_i$, and then compute $\sum_{u \colon r_{ui} = 0} \vf_u$ via 
\[
\sum_{u \colon r_{ui} = 0} \vf_u = \sum_u \vf_u  - \sum_{u \colon r_{ui} = 1} \vf_u. 
\]
This idea is taken from \cite{bigclam}, where it was used in the context of community detection. 
Using the precomputed $\sum_{u \colon r_{ui} = 0} \vf_u$, a gradient descent step of updating $\vf_i$ has cost $O(|\{ u \colon r_{ui}=1\}|   K )$. 
Thus, updating all $\vf_i$ and all $\vf_u$ has cost $O(|\{(i,u)\colon r_{ui}=1\}|  K )$, which means that updating all factors has a cost that is linear in the problem size (i.e., number of positive examples) and linear in the number of co-clusters.

\subsection{Relation to Matrix Factorization}
\label{sec:mtxfact}

Latent factor models are a prevalent approach to recommender systems. Among them, matrix factor\-ization approaches are particularly popular, because of their
good performance properties. 
In this section, we discuss the connection of our approach to standard matrix factorization techniques. 
Matrix factorization models in their most basic form represent users and items by latent vectors $\vf_u, \vf_i$ in a low-dimensional space $\reals^K$, and form a rating according to 
\[
r_{ui} = \innerprod{\vf_u}{\vf_i}. 
\]
A common approach to fit the latent factors based on a given set of examples $\{r_{ui}\}$ is to 
\begin{align}
\text{minimize } \sum_{u,i} \ell(r_{ui},  \innerprod{\vf_i}{\vf_u}) + \lambda \norm[2]{\vf_i}^2 + \lambda \norm[2]{\vf_u}^2,
\label{eq:stdmf}
\end{align}
where $\ell$ is a loss function and $\norm[2]{\vf_i}^2, \norm[2]{\vf_u}^2$ are regularization terms to prevent overfitting. 
A common choice for non-binary ratings is the quadratic loss function $\ell(r_{ui},  \innerprod{\vf_i}{\vf_u}) =  (r_{ui} - \innerprod{\vf_i}{\vf_u})^2$. 
For the one-class collaborative filtering problem, quadratic loss is not directly applicable, as the question remains on how to deal with the unknowns, for which $r_{ui}=0$. 
Performing summation in \eqref{eq:stdmf} only over the positive examples ($r_{ui}=1$) is not sensible, as it results in a trivial solution of $\vf_u = \vf_i$, for all $u$ and all $i$ to \eqref{eq:stdmf}. A different approach is to treat the unknowns as negative ratings. However, this might bias recommendations as some of the unknown examples are actually positive ratings. 
 To resolve this issue, it has been proposed in \cite{pan_one-class_2008} to give different weights to the error terms corresponding to positive and unknown ratings in the objective function, specifically to use the cost function
\begin{align}
\ell(r_{ui}, \innerprod{\vf_i}{\vf_u}) 
=
\begin{cases}
(r_{ui}  -  \innerprod{\vf_i}{\vf_u} )^2 & \text{ if } r_{ui} = 1 \\
b (r_{ui}  -  \innerprod{\vf_i}{\vf_u} )^2 & \text{ if } r_{ui} = 0
\end{cases},
\label{eq:costfuncwALS}
\end{align}
where $b<1$ is a weight assigned to unknown ratings. A common approach to solve the corresponding optimization problem is weighted alternating least squares (wALS) \cite{pan_one-class_2008}. 

Our approach is equivalent to choosing the loss function 
\begin{align}
\ell(r_{ui}, \innerprod{\vf_i}{\vf_u}) 
=
-\log( | r_{ui} - e^{- \innerprod{\vf_i}{\vf_u}} |).
\label{eq:losfunct}
\end{align}
This results in a large penalty if $\innerprod{\vf_i}{\vf_u}$ is small for a positive example $r_{ui}=1$, and a moderate penalty if $\innerprod{\vf_i}{\vf_u}$ is large for an unknown example $r_{ui}=0$. Our approach is therefore similar in spirit to that of giving different weights to positive ($r_{ui}=1$) and unknown ($r_{ui}=0$) examples. 

\medskip
\noindent\textbf{Interpretability:}
Matrix factorization approaches, such as the wALS algorithm (with loss function \eqref{eq:costfuncwALS}), yield good empirical performance, as we will see in the experiments. Their main disadvantage for our scenario is that the latent space is typically not easy to interpret. This drawback of MF techniques is attested in various studies \cite{NMF,nonInterpretable1, nonInterpretable2, nonInterpretable3}. %, and in fact, latent factors are hardly interpretable. 
This statement also applies to standard non-negative matrix factorization (NMF) techniques, where the factors are constraint to be non-negative. 
Our approach also uses factorization principles, however, we \textit{confine the factors to explicitly model user and item participation}, which is key for not compromising the model interpretability.

%%%%%%%%%%%%%%%%%%%%%%%%%%%%%%%%%%%%%%%%%%%%%%%%%%%%%%%%%%%%%%%%%%%%%%%%%%%%%%%%%%%%%%%%

%%%
% suppress if needed
%\pagebreak
%%%%%%%%%%%%%%%%%%%%%%%%%%%%%%%%%%%%%%%%%%
\section{Relative OCuLaR algorithm (R-OCuLaR)}
\label{sec:relativeOCuLaR}
As we saw in the related work section, OCCF problems can be viewed as learning of either absolute or relative preferences of users. The OCuLaR algorithm belongs to the first category because it tries to learn the absolute ratings of the users.
Specifically, as we saw, OCuLaR can be viewed as a non-negative matrix factorization approach with a particular loss function (i.e., \eqref{eq:losfunct}). The loss function assigns a large penalty to 
positive examples that are not well explained by the model (i.e., $\innerprod{\vf_i}{\vf_u}$ small for $r_{ui}=1$), and only a moderate penalty to unknown examples that are not well explained by the model (i.e., $\innerprod{\vf_i}{\vf_u}$ large for $r_{ui}=0$). 
%In spirit, OCuLaR is therefore similar to the general approach of assigning different weights to positive and negative examples. 

%The OCuLaR algorithm therefore tries to fit the \emph{absolute} values of the ratings. 

In this section, we examine how OCuLaR could be adapted to treat the positive examples as \textit{relative} preferences.
The notion of relative preferences in OCCF was first explored by Rendle et al.~\cite{Rendle2009}, who proposed to predict the relative preferences ($u$ prefers $i$ over $j$) rather than the absolute rankings. 
The underlying idea for this case is that there is an associated latent personalized ranking $>_u$ for each user $u$, where $i>_u j$ signifies that user $u$ prefers item $i$ over item $j$. 
To this end, in a first step, one can construct a ``training'' item ranking set for each user from the set of positive examples, denoted by $S = \{(u,i) \colon r_{ui}=1\}$. 
The underlying assumption for this constuction is that, if $r_{ui}=1$, then user $u$ prefers item $i$ item over all items $j$ with unknown rating, i.e., $r_{uj} = 0$. 
%For item pairs $(i,j)$ with $r_{ui}=1$ and $r_{uj} = 1$, or with $r_{ui}=0$ and $r_{uj} = 0$, we can obviously not obtain any preference. 
Specifically, the training item ranking data set $D_S$ is defined as:
\[
D_S = \{(u,i,j) \colon r_{ui}=1 \text{ and } r_{uj}=0 \}. 
\]
%contains all triplets $(u,i,j)$, which correspond to the items $i$ being preferred over item $j$ by user $u$. 
Note that negative examples are accounted for implicitly because if $(u,i,j) \in D_S$, then $j$ is not prefered over $i$. 
Triplets not in $D_S$ correspond to triplets for which no direct preference information is available; it is those triplets for which we want to learn the preferences. 

As shown in \cite{Rendle2009}, assuming that the users act independently of each other and that $>_u$ is independent across $u$, 
the model likelihood can be maximized by maximizing
\begin{align}
\prod_{(u,i,j) \in D_S}   
\PR{i >_u j }. 
\label{eq:proxylogli}
\end{align}
To adapt the notion of relative preferences for OCuLaR, suppose that the probability of a user preferring item $i$ over item $j$ is given by
\[
\PR{i >_u j} = (1 - e^{-\innerprod{\vf_u}{\vf_i}}) e^{-\innerprod{\vf_u}{\vf_j}}. 
\]
To see the formal relation to the model in Section \ref{seg:genmod}, simply note that, with this choice,
\[
\PR{i >_u j} = \PR{r_{ui}=1} \PR{r_{uj}=0},
\]
with $\PR{r_{ui}=1}$ as defined in \eqref{eq:prrui1}. 
Maximizing \eqref{eq:proxylogli} is equivalent to minimizing the logarithm of \eqref{eq:proxylogli}, i.e., minimizing
%\vspace{-\baselineskip}
%\begin{align}
%&\sum_{(u,i,j) \in D_S} 
%\left(
%\log (1 - e^{-\innerprod{\vf_u}{\vf_i}})
%+\log e^{-\innerprod{\vf_u}{\vf_j}} \right) \nonumber \\
%
%&\propto 
%\sum_{(u,i) \colon r_{ui} = 1}
%w_u
%\log (1 - e^{- \innerprod{\vf_u}{ \vf_i } } )
%-
%\sum_{(u,j)\colon r_{uj} = 0} \innerprod{\vf_u}{ \vf_i } 
%\end{align}
\begin{align}
&\prod_{(u,i,j) \in D_S}   
\PR{i >_u j | \theta} \nonumber \\
&=
\sum_{(u,i,j) \in D_S} 
\left(
\log \PR{ r_{ui} = 1 }
+\log \PR{ r_{uj} = 0 } \right) \nonumber \\
&= 
\sum_{u}
\sum_{ i \colon r_{ui} = 1} \sum_{j\colon r_{uj} = 0}
\left(
\log \PR{ r_{ui} = 1 }
+\log \PR{ r_{uj} = 0 } 
\right) \nonumber
%\nonumber \\
%
\end{align}
\begin{align}
&= 
\sum_{u} 
|\{j \colon  r_{uj} = 0 \}|
\sum_{i \colon r_{ui} = 1}
\log \PR{ r_{ui} = 1 } \nonumber\\
&+
| \{ i \colon r_{ui} = 1 \} | \sum_{j\colon r_{uj} = 0}
\log \PR{ r_{uj} = 0 } \nonumber \\
&\propto 
\sum_{u} 
w_u
\sum_{i \colon r_{ui} = 1}
\log \PR{ r_{ui} = 1 }
+
\sum_{j\colon r_{uj} = 0}
\log \PR{ r_{uj} = 0 } \nonumber \\
&=
\sum_{(u,i) \colon r_{ui} = 1}
w_u
\log (1 - e^{- \innerprod{\vf_u}{ \vf_i } } )
-
\sum_{(u,j)\colon r_{uj} = 0} \innerprod{\vf_u}{ \vf_i }, \nonumber
\end{align}
where we defined $w_u = \frac{|\{ i \colon r_{ui} = 0 \}|}{ |\{ i \colon r_{ui} = 1 \}| }$. 
Note that the RHS above  differs only in the factors $w_u$ from the negative log-likelihood \eqref{eq:modlogli} in Section \ref{sec:fittinmp}. 
The factors $w_u$ assign a large weight to \emph{positive} examples of users that have few positive examples associated with them. 

We denote the algorithm obtained by substituting the log-likelihood $-\log \mc L$ in Section \ref{sec:fittinmp} with the RHS above as the \textit{relative OCuLaR (R-OCuLaR) algorithm}. Its implementation is essentially equivalent to that of the (original) OCuLaR algorithm; in fact, it has exactly the same complexity.

\section{Using Massively Parallel Processors}
\label{sec:gpu}

The OCuLaR algorithm can be dramatically accelerated by exploiting its inherent parallelism. In fact, one can easily map and implement the costly training phase onto a graphics processing unit (GPU). This results in two benefits:

\begin{itemize}
    \item Leverage the \textbf{reduction in training time} to train the model and produce recommendations in significantly reduced time. For the real industrial dataset used in our experiments, both training of the model and generation of recommendations can be completed in mere seconds when using a GPU.
    \item Perform a more \textbf{fine-grained grid search} over the algorithm's hyper-parameters to obtain a higher precision/recall. OCuLaR requires the learning of $K$ (number of co-clusters) and $\lambda$ by using a cross-validated grid search. This phase can be costly, so in practice one can only search across a certain range of values and keep the pair that results in the best performance under the desirable accuracy metric. Achieving very fast learning using GPUs essentially allows the methodology to examine a broader range of hyper-parameter value pairs that help improve the overall accuracy.
\end{itemize}
 In order to achieve the potential acceleration offered by GPU devices, communication of large amounts of data between the host and the device must be avoided, especially during iterative computations. Furthermore, arithmetic computations must be carefully designed to optimally make use of the underlying computational architecture and memory hierarchy. Below, we describe in more detail how OCuLaR can be mapped onto an efficient GPU implementation.

\subsection{GPU implementation}

We begin by describing how data moves between the host and the GPU during the training of the OCuLaR algorithm. Firstly, the entire training data (in a sparse format) is copied from the host memory into the GPU main memory along with a set of initial values for the $\vf_i$ and $\vf_u$ vectors. The iterative training algorithm described in Section \ref{sec:impdet} is then performed by launching a sequence of kernel functions. Throughout the training, all data remains on the GPU and communication back and forth between the host and the device is limited to a small amount of control logic. Once a predetermined set of iterations has been executed, the learned values of the $\vf_i$ and $\vf_u$ are copied back from the GPU to the host memory and the training is completed. 

%\vspace{-\baselineskip}
\begin{figure}[!h]
\centering
\includegraphics[width=\columnwidth]{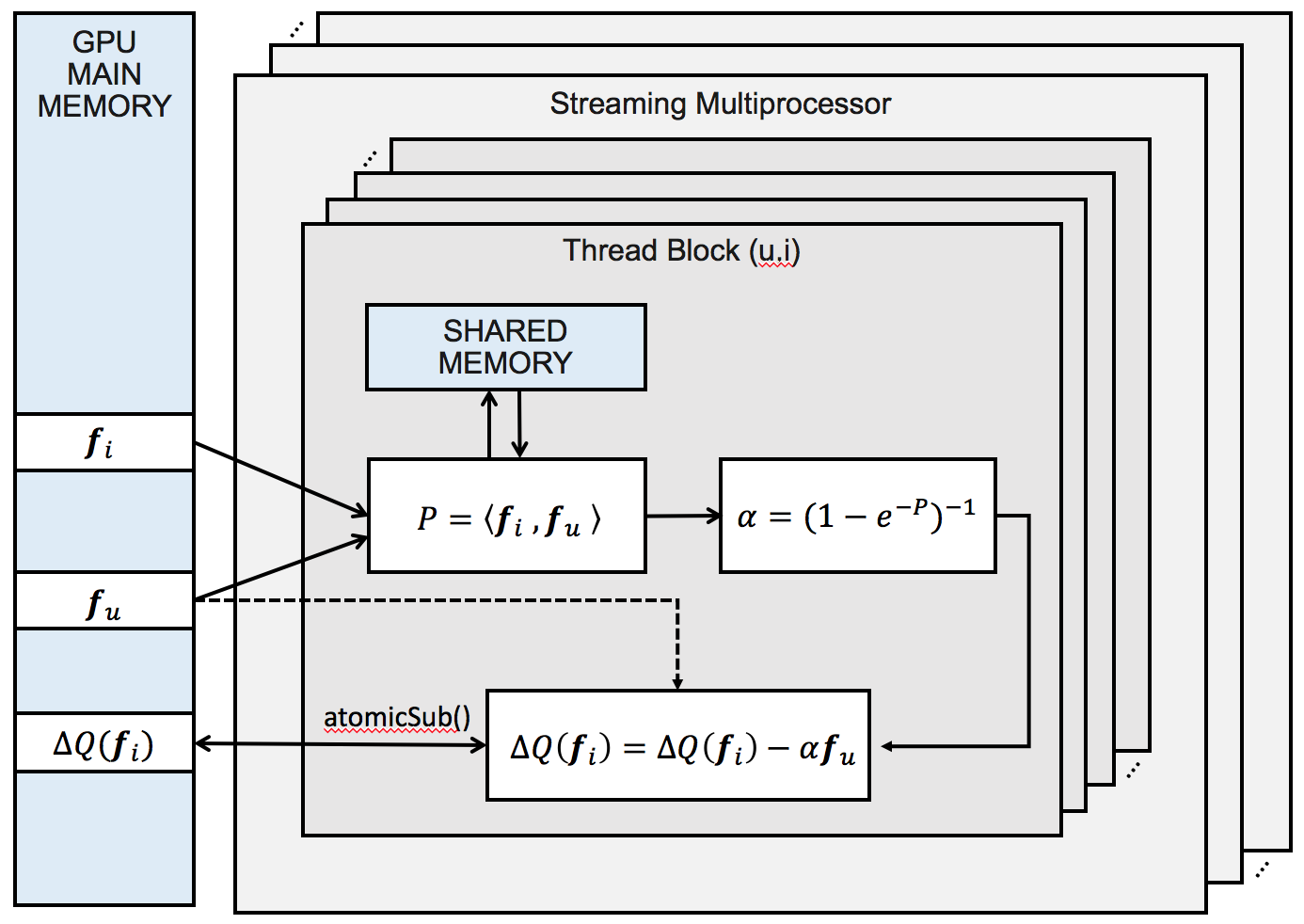}
\caption{Computation of the gradient vector is performed by a large number of thread blocks that are executed on the streaming multiprocessors of the GPU}
\label{fig:gpu_arch}
\end{figure}

The arithmetic computations that comprise the training are split into a number of separate kernel functions. As an illustration, we provide a detailed description of how the kernel that computes the gradient vectors for all items is structured to efficiently use the underlying hardware. We begin by noting that equation (\ref{eq:origgrad}) can be expressed:
\begin{eqnarray}
\nabla Q( \vf_i ) 
&=& \sum_{u}\vf_u + 2 \lambda \vf_i - 
\sum_{u\colon r_{ui} = 1}  \vf_u \left(1 -e^{-  \innerprod{\vf_u}{ \vf_i} }\right)^{-1}\nonumber\\
&=& C + 2 \lambda \vf_i - \sum_{u\colon r_{ui} = 1}  \vf_u \alpha(\innerprod{\vf_u}{ \vf_i}),
\label{eq:gpugrad}
\end{eqnarray}
where $C$ is a constant independent of the item index. As in the reference CPU-only implementation, an initial computation is performed to calculate the sum of all $\vf_u$ vectors and the gradient vectors are initialized in GPU memory as $C+2\lambda\vf_i$. A kernel function is then called which launches a thread block for every positive rating in the training data. Since this number is typically very large, this mapping is well suited to the massively parallel GPU architecture in which a number of streaming multiprocessors are capable of executing multiple thread blocks concurrently. 

We will now describe the computation performed by a single thread block in more detail, making reference to Figure \ref{fig:gpu_arch}. The thread block first fetches the corresponding $\vf_i$ vector and $\vf_u$ vector from GPU main memory and computes the inner product between the two. Following the approach of \cite[p. 76]{sanders2010cuda}, an individual thread within the block handles computation of only part of the inner product. The memory access patterns are carefully arranged so that threads that are executed concurrently (as a "warp") access contiguous regions of memory, allowing reads to be effectively coalesced. The partial results are then stored in high-speed shared memory and a reduction is performed to obtain the final value of the inner product ($P$). A single thread within the block then computes the scalar $\alpha$. Finally, the thread block multiplies this scalar by the $\vf_u$ vector and updates the corresponding item gradient in GPU main memory using atomic operations. In this manner, the sum in equation~\eqref{eq:gpugrad} is computed entirely asynchronously for all items. Once all the thread blocks have finished execution, the correct value of the gradient vectors exist in GPU main memory and the training can proceed. The other essential computations, such as the evaluation of the likelihood function are mapped to the GPU in a similar fashion.

\noindent\textbf{Memory:} The memory footprint of the GPU-based OCuLaR implementation scales as:
\begin{equation}
O\left(\max
\left(|\{(u,i):r_{ui}=1\}|,n_uK,n_iK\right)\right),\nonumber
\end{equation}
where $n_u$ and $n_i$ are the number of users and the number of items respectively. This property allows for training on very large datasets despite the relatively limited memory capacity of modern GPU devices. For example, around 2.7GB of GPU memory is required to train on the Netflix dataset (assuming $K=200$) and thus the problem easily fits within the main memory of an inexpensive GPU (typically up to 12GB). In contrast, a previous attempt to implement an alternating-least-square based matrix factorization approach on GPUs determined that the memory requirements for the same dataset exceeded 12GB (for the equivalent of K=100) \cite{tan2016faster}.  

\ignore{
%%%
\medskip\noindent
\textbf{Implementation and complexity:}
The user and item latent factors $\vf_u, \vf_i$ are updated as before using a projected gradient approach. The only difference is that we have to add an extra term to $Q(\vf_i)$ and its gradient $\nabla Q(\vf_i)$ in \eqref{eq:coordstep} and \eqref{eq:origgrad}, respectively.

Specifically, setting for convenience $\vf'_i = [\vf_i; 1]$ and $\vw'_\ell = [\vw_\ell; w_{0\ell}]$, the attribute log-likelihood $-\log \mc L_A$ becomes
\begin{align}
%%
%&=
%\sum_{i\colon A_i = a_k, k\neq J-1}
%-\log \PR{A_{i} = a_{k} | \mF}
%+
%\sum_{i\colon A_i = a_{J-1}}
%-\log \PR{A_{i} = a_{J-1} | \mF} \nonumber \\
%%
%&=
%\sum_{i\colon A_i = a_k, k\neq J-1}
%-\log \frac{ 
%e^{- \innerprod{\vw'_k}{ \vf'_i }  }
%}{   
%1 + \sum_{\ell=0}^{J-2}
%e^{- \innerprod{\vw'_\ell}{ \vf'_i } }
%}
%+
%\sum_{i\colon A_i = a_{J-1} }
%-\log
%\frac{ 1}{   
%1 + \sum_{\ell=0}^{J-2} e^{- \innerprod{\vw'_\ell}{\vf'_i } }
%} \nonumber \\
%
\sum_i \log\left( 1 + \sum_{\ell=0}^{J-2} e^{- \innerprod{\vw'_\ell}{ \vf'_i } }\right)
+
\sum_{i\colon A_i = a_k, k\neq J-1}    \innerprod{\vw'_k}{ \vf'_i }. \nonumber
\end{align}
The part of $-\log \mc L_A $ depending on $\vf_i$ that we add to \eqref{eq:coordstep} is given by
\begin{align}
\log
\left(
1 + \sum_{\ell=0}^{J-2} e^{- \innerprod{\vw'_\ell}{\vf'_i } } 
\right) 
+
\ind{A_i = a_k}
\innerprod{\vw'_k}{ \vf'_i }
\label{eq:part1}
\end{align}
and the corresponding gradient we need to add to \eqref{eq:origgrad} is given by 
\begin{align}
%\nabla(- \log \mc L_A(\vf_i)  )
%\frac{\partial ( -\log \mc L_A)}{\partial \vf_i}
%
%&=
- 
\left( 1 + \sum_{\ell=0}^{J-2} e^{- \innerprod{\vw'_\ell}{ \vf'_i } }\right)^{-1}
\sum_{\ell=0}^{J-2} e^{- \innerprod{\vw'_\ell}{ \vf'_i } }  \vw_\ell
+
%\sum_{i\colon A_i = a_k, k\neq K}   
\ind{A_i = a_k}
\vw_k .
\label{eq:part2}
\end{align}
Thus the complexity to update all the factors $\vf_i,\vf_u$ is
\[
O(|\{(i,u)\colon r_{ui}=1\}|  K  +  J K N),
\]
where $N$ is the number of items.  

\medskip\noindent
\textbf{Updating the weights:}
To update the weights $\vw'_\ell$, we use gradient descent. The part of  $\tilde Q$ depending on $\vw'_\ell$ is
\[
\tilde Q(\vw'_\ell)
=
\sum_i \log\left( 
1
+
%\sum_{\ell=1}^{J-1} e^{- \innerprod{\vw'_\ell}{ \vf'_i } }
\sum_{j = 0}^{J-2} e^{- \innerprod{\vw'_j}{ \vf'_i } }
%e^{- \innerprod{\vw'_\ell}{ \vf'_i } }
%+
%1 
%+ 
%%\sum_{\ell=1}^{J-1} e^{- \innerprod{\vw'_\ell}{ \vf'_i } }
%\sum_{j = 0, j \neq \ell}^{J-2} e^{- \innerprod{\vw'_j}{ \vf'_i } }
\right)
+
\sum_{i\colon A_i = a_\ell }    \innerprod{\vw'_\ell}{ \vf'_i },
\]
and its gradient is
\begin{align*}
&\nabla \tilde Q(\vw'_\ell)  \nonumber  \\
%\frac{\partial \tilde Q(\vw'_\ell) }{\partial \vw'_\ell}
%=
%\frac{\partial ( -\log \mc L_A)}{\partial \vw'_\ell}
%
&=
- 
\sum_{ i } 
\left( 1 + \sum_{j=0}^{J-2} e^{- \innerprod{\vw'_j}{ \vf'_i } }\right)^{-1}
e^{- \innerprod{\vw'_\ell}{ \vf'_i } }  \vf'_i
+
\sum_{i\colon A_i = a_\ell}  
\vf'_i .
\end{align*}
Both $\tilde Q(\vw'_\ell)$ and the gradient $\nabla \tilde Q(\vw'_\ell)$ depend on the sum $\sum_{j=0}^{J-2} e^{- \innerprod{\vw'_j}{ \vf'_i }}$. To speed up computation, we therefore pre-compute $\sum_{j=0}^{J-2} e^{- \innerprod{\vw'_j}{ \vf'_i }}$ before updating the weights $\vw_\ell'$. 
We then update the corresponding sum after having updated $\vw'_j$. 
With this pre-computation step, the complexity of updating all weights is $O(J K N)$.  
}

%%%%%%
\section{Experiments}
\label{sec:exper}

Here, we compare the prediction accuracy of OCuLaR to that of other interpretable and non-interpretable one-class CF algorithms.
We compare with interpretable user-based and item-based collaborative filtering approaches,
and non-inter\-pre\-table state-of-art one-class recommendation algorithms based on  matrix factorization. 
Both OCuLaR and R-OCuLaR typically outperform or are on par with the existing one-class recommendation algorithms in terms of recommendation performance.
We also discuss the choice of the input parameters for our approach. We show that OCuLaR exhibits linear scalability,  demonstrate the acceleration achieved by the GPU implementation, and conclude with an industrial deployment.

\subsection{Datasets}
We use four datasets.
First, we consider a real-world dataset from our institution. It consists
of the buying history of 80,000 clients with whom our institution interacts and 3,000 products or services offered by our institution. The clients in this case are not individuals but companies, and the recommender system operates in a B2B setting. We call this dataset \textit{B2B-DB}.
%The dataset is highly imbalanced in the sense that the most frequently purchased item was bought by more than $50\%$ of the users, whereas the most infrequently purchased item was bought only $50$ times. 
We also consider three public datasets.
The first was extracted from the \textit{CiteULike} website, which assists users in creating collections of scientific articles \cite{wang_collaborative_2015}. The dataset consists of 5,551 users and 16,980 articles. Each user has a number of articles in their collection, which are considered as positive examples. Based on the positive examples, the goal is to generate new article recommendations. 
The second dataset is the \textit{Movielens} 1 million dataset, which consists of 1 million ratings from 6,000 users on 4,000 movies. 
The final dataset is the \textit{Netflix} dataset, consisting of about 100 millions ratings that 480,189 users gave to 17,770 movies. 
%To compare  OCuLaR with other approaches we use a smaller instance of Netflix dataset consisting of 5,000 random users and 5,000 random items (Netflix5K). This is because some of the other techniques are not applicable for very large datasets. We use the full Netflix data in the scalability experiment of OCuLaR.

In both the Movielens and the Netflix dataset, the users provide ratings between 1 and 5 stars. As we consider a one-class collaborative filtering task, we adopt the convention from many previous works (e.g., \cite{GBPR, NMF1class}) to
only consider ratings greater than or equal to $3$ as positive examples and ignore all other ratings. Therefore, the task now is equivalent to predicting whether the user will give a rating greater than 3 (i.e., is likely to enjoy the movie).

%%%%%%%%%%%%%%%%%%%%%%%%%%%%%%%%%%%%%%%%%%%%%%%%%%%%%%%%%%%%%%%%%%%%%%%%%
\begin{table*}[!ht]
\centering
\caption{Comparison of OCuLaR and R-OCuLaR with other baseline one-class recommendation algorithms.}
\label{tab:results}
\footnotesize{
\begin{tabular}{l l c c c c c c } 
\hline
 & & & & & &  \textbf{user-}& \textbf{item-} \\
 dataset & metric & \textbf{OCuLaR} & \textbf{R-OCuLaR} &  \textbf{wALS} & \textbf{BPR} & \textbf{based} & \textbf{based}\\ \hline
 %&  &  &  &  \cite{pan_one-class_2008, Koren2008} &  \cite{Rendle2009} & \cite{sarwar_analysis_2000} & \cite{deshpande_item-based_2004}\\ \hline
 \multirow{2}{*}{Movielens}  & MAP@50 & \textbf{.1809} & .1805 & .1513 & .1434 & .1639 & .1329  \\ 
                            & recall@50   & .4021 & \textbf{.4086} &  .3982 & .3587 & .3757 & .3238 \\ \cline{1-8}
                            \hline
\multirow{2}{*}{CiteULike}  & MAP@50 & .0906 & .0916 & .1003 & .0157 & .0882 & \textbf{.1287}  \\ 
                            & recall@50   & .3042 & .3177 &  \textbf{.3331} & .0801 &  .2699 & .2921 \\ \cline{1-8}
                            \hline
\multirow{2}{*}{B2B-DB} & MAP@50 & \textbf{.1801} & .1651 & .1749 & .1325 & .1797 & .1568  \\ 
                            & recall@50   & .5240 & .4780 &  \textbf{.5283} & .4407 & .4995 & .4840 \\ \cline{1-8}
\end{tabular}
}

%\vspace{-0.5cm}
\end{table*}

%%%%%%%%%%%%%%%%%%%%%%%%%%%%%%%%%%%%%%%%%%%%%%%%%%%%%%%%%%%%%%%%%%%%%%%%%

\subsection{Recommendation performance}

\subsubsection{Evaluation Metrics}

%\RH{I would like to be precise here. A reader who understands the metrices will not look at this part in any case, and a reader who does not fully understand the metrices, needs a precise definition. Also the previous text contained an exact copy of a sentence from the bookd ``Introduction to information retrieval'', and was therefore problematic. }

We measure performance in terms of \emph{recall at $M$ items} (recall@$M$), and \emph{mean average precision at $M$ items} (MAP@M).

In the one-class setting, recall is a more sensible measure than precision, because an example being unknown ($r_{ui}=0$) does not mean that user $u$ would not rate item $i$ positively  \cite{DBLP:conf/sigir/ScheinPUP02}. 
%Because an example $r_{ui}$ not being positive does not mean that $u$ is not interested in item $i$, an accurate comparison of the precision is difficult.  However, recall is a sensible performance criterion. 
Given an ordered (by relevance) list of $M$ recommendations for user $u$, denoted by $i_1,...,i_M$, the recall@M items for user $u$ is defined as 
\[
\begin{small}
\text{recall}@M(u)
=
\frac{ |\{i\colon r_{ui} =1 \} \cap \{i_1,...,i_M\} |}{ |\{i\colon r_{ui}=1\}| }.
%%\frac{\text{number of articles user $u$ bought of the set of $M$ recommendations}}{\text{total number of buys of user $u$}}
\end{small}
\]
The overall $\text{recall}@M$ is obtained as the average over $\text{recall}@M(u)$. 

MAP is commonly used in information retrieval for evaluating a ranked or ordered list of items, and is considered a good measure of performance when a short list of the most relevant items is shown to a user. 
 MAP@M items is the mean (over all users) of the average precision at $M$ items (AP@M), defined as
\[
\text{AP}@M(u)  
= 
\sum_{m=1}^M \text{Prec}(m)  
\frac{  \ind{ r_{ui_m} = 1 } }{\min( |\{i\colon r_{ui}=1\}| , M)},
\]
where $|\{i\colon r_{ui}=1\}|$ is the number of positive examples corresponding to user $u$; $\ind{ r_{ui_m} = 1}$ is equal to $1$ if $r_{ui_m}=1$ and zero otherwise, 
and $\text{Prec}(m)$ is the precision at a cutoff rate $m$:
\[
\text{Prec}(m) = \frac{ |\{i\colon r_{ui} =1 \} \cap \{i_1,...,i_m\}|}{ m }. 
\]
%Note that $\frac{\delta(m)}{\min(N , M)}$ can be understood as the change in recall.
Note that because $\text{Prec}(m) \in [0,1]$, we have that $\text{AP}@M \leq 1$. 
%
%%``MAP provides a single-figure measure of quality across recall levels, has especially good discrimina- tion and stability properties, and roughly corresponds to the average area under the precision-recall curve'' -> C. D. Manning, P. Raghavan, and H. Schu?tze. Introduction to information retrieval. 
%%Let $M$ be the number of recommendations given, and let $N$ be the number of true relevant items. The average precision at $M$ is given by
%%\[
%%\sum_{m=1}^M \text{Prec}(i)  \text{Change in recall at $m$}.
%%\]
%%the change in recall at $i$ is $1/M$ if $m$ is correct, otherwise $0$. If the number of relevant items $N$ is smaller than $M$, then the change in recall is $1/N$. 

%%%%
\subsubsection{Comparison with baselines}

We compare the OCuLaR algorithm with various interpretable and non-interpretable baseline algorithms:

\begin{itemize} 
 \item \textbf{User-based} collaborative filtering using a cosine similarity metric (e.g., \cite{sarwar_analysis_2000}). Such an algorithm is interpretable because a recommendation can be justified with a reasoning of the type:  ``item $i$ is recommended because the similar users $u_1,..,u_k$ also bought item $i$''. 

 \item\textbf{Item-based} collaborative filtering using cosine similarity (e.g., \cite{deshpande_item-based_2004}). This algorithm is also interpretable. It can be accompanied with a recommendation such as: ``item $i$ is recommended because user $u$ bought the similar items $i_1,...,i_k$''. 
%\item  \textbf{1c-NMF} A one-class non-negative matrix factorization approach \cite{NMF1class}. 

\item Weighted Alternating Least Squares (\textbf{wALS}), a state-of-the-art one-class matrix factorization approach \cite{pan_one-class_2008}.  %Koren2008} 
wALS minimizes the loss function by alternatingly optimizing the user and the item factor via least-squares.
%wALS minimizes \eqref{eq:stdmf} with loss function \eqref{eq:costfuncwALS} by alternatingly optimizing the user and the item factor via least-squares. 
It offers good prediction performance but the reasoning is not directly interpretable, as mentioned earlier.

\item %\textbf{GBPR}. Group bayesian personalized ranking (GBPR) represents the state-of-the-art optimization framework of CF for binary relevance data \cite{GBPR}. GBPR was shown to outperform the previous state-of-art approach, the BPR approach by Rendle et al. \cite{Rendle2009}. For GBPR we used the code from the publicly available recommendation suite LibRec \cite{LibRec}. 
Bayesian personalized ranking (\textbf{BPR}), a state-of-the-art matrix factorization approach that converts the set of positive examples into a set of relative preferences \cite{Rendle2009} (see also Section \ref{sec:relativeOCuLaR}). The recommendations provided by BPR are also not directly interpretable. 
For BPR we used the python/theano implementation from \url{https://github.com/bbcrd/theano-bpr.git}. 
\end{itemize}

All approaches require the setting of hyper-parameters.
For each technique we test a number of hyper-parameters and report only the best results. 
%Specifically, for OCuLaR and R-OCuLaR, we performed a grid search over $(K,\lambda) \in \{60,80,100,120,140\} \times \{60,80,100,\allowbreak 120,140\}$. For the User- and Item-based collaborative filtering approaches we selected the number of nearest-neighbors in $k\in \{40,50,60,70,\allowbreak80,90,100\}$ and $k\in\{15,25,35,45,65,\allowbreak85\}$ respectively. For wALS we choose the weight in the loss function \eqref{eq:costfuncwALS} as $b=0.01$, the regularization parameter $\lambda=0.01$, and consider the choices $K \in \{40,50,60,\allowbreak70,80\}$ for the dimension of the latent vectors. Finally, for BPR we consider the following values for the dimension of the latent factors, $K$, and the regularization parameter, $\lambda$: $(K,\lambda) \in \{10,20,30,40\} \times \{0.0005,0.0015,\allowbreak0.0025,0.005,\allowbreak0.01,0.05,\allowbreak0.01\}$. 
 For OCuLaR and R-OCuLaR%\footnote{\scriptsize Source code at \url{http://alumni.cs.ucr.edu/~mvlachos/erc/projects/recommendations/}}
, we performed a grid search over $K$ and $\lambda$ (within the range 100 and 200). As we show later, this range does not include the parameters leading to the `optimal' results, but it is sufficient to showcase a baseline performance without excessive tuning. For the user- and item-based collaborative filtering approaches, we performed a grid search over the number of nearest neighbors. For wALS, we choose the weight in the loss function \eqref{eq:costfuncwALS} as $b=0.01$, the regularization parameter as $\lambda=0.01$, and performed a grid search over the dimension of the latent vectors. Finally, for BPR we searched over the dimension of the latent factors $K$, and the regularization parameter $\lambda$. 
We computed the recall@M and MAP@M by splitting the datasets into a training and a test dataset, with a splitting ratio of training/test of 75/25, and averaging over 10 problem instances.

%%%%%%%%%%%%%%%%%%%%%%%%%%%%%%
\vspace{-\baselineskip}
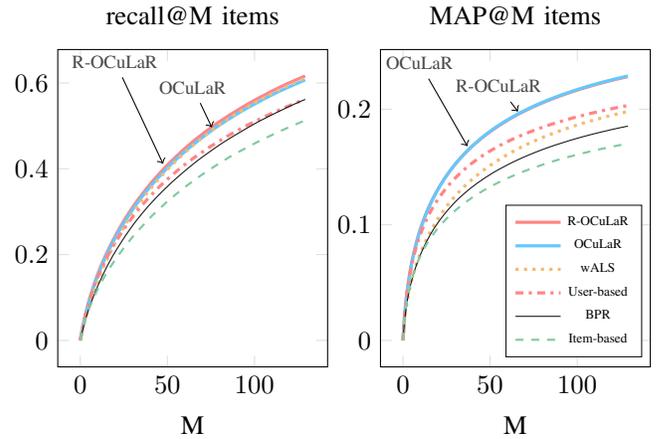
\begin{figure}[!ht]
\centering
\begin{tikzpicture}    

\begin{groupplot}[
width=0.285\textwidth,
height = 0.32\textwidth,
group style={group size=2 by 1,
xticklabels at=edge bottom, xlabels at=edge bottom,
%yticklabels at=edge left, ylabels at=edge left,
horizontal sep=20pt,
vertical sep=4pt,
},
 every axis y label/.style=
            {at={(ticklabel cs:0.5)},rotate=90,anchor=near ticklabel},
xlabel = M,
]

\nextgroupplot[title = recall@M items, legend style={legend pos=south east,}]
%    \begin{axis}[
%            xlabel=M,
%            xmax=40,
%    ylabel=recall@M items,
%    width =0.7\textwidth,
%    legend style={legend pos=south east,},
%    every axis plot post/.append style= {mark=none},]

\addplot +[] table[x index=0,y index=4]{./data/ML1m_gridsearch_recall.dat};
%\addlegendentry{R-OCuLaR};
\addplot +[] table[x index=0,y index=3]{./data/ML1m_gridsearch_recall.dat};
%\addlegendentry{OCuLaR};
\addplot +[] table[x index=0,y index=5]{./data/ML1m_gridsearch_recall.dat};
%\addlegendentry{wALS};
\addplot +[] table[x index=0,y index=2]{./data/ML1m_gridsearch_recall.dat};
%\addlegendentry{User KNN};
\addplot +[] table[x index=0,y index=6]{./data/ML1m_gridsearch_recall.dat};
%\addlegendentry{BPR};
\addplot +[] table[x index=0,y index=1]{./data/ML1m_gridsearch_recall.dat};
%\addlegendentry{Item KNN};

\node[anchor=west] (source) at (axis cs:-10,0.65){\scriptsize \textcolor{textgray}{R-OCuLaR}};
       \node (destination) at (axis cs:50,0.39){};
       \draw[->](source)--(destination);
       
\node[anchor=west] (source) at (axis cs:40,0.59){\scriptsize \textcolor{textgray}{OCuLaR}};
       \node (destination) at (axis cs:78,0.475){};
       \draw[->](source)--(destination);   

%%%%%%%%%%%%
 
\nextgroupplot[title = MAP@M items,legend style={legend pos=south east,}]
\addplot +[] table[x index=0,y index=4]{./data/ML1m_gridsearch_MAP.dat};
\addlegendentry{\tiny R-OCuLaR};

\addplot +[] table[x index=0,y index=3]{./data/ML1m_gridsearch_MAP.dat};
\addlegendentry{\tiny OCuLaR};

\addplot +[] table[x index=0,y index=5]{./data/ML1m_gridsearch_MAP.dat};
\addlegendentry{\tiny wALS};

\addplot +[] table[x index=0,y index=2]{./data/ML1m_gridsearch_MAP.dat};
\addlegendentry{\tiny User-based};

\addplot +[] table[x index=0,y index=6]{./data/ML1m_gridsearch_MAP.dat};
\addlegendentry{\tiny BPR};

\addplot +[] table[x index=0,y 
index=1]{./data/ML1m_gridsearch_MAP.dat};
\addlegendentry{\tiny Item-based};

\node[anchor=west] (source) at (axis cs:-15,0.24){\scriptsize \textcolor{textgray}{OCuLaR}};
       \node (destination) at (axis cs:40,0.16){};
       \draw[->](source)--(destination);
       
\node[anchor=west] (source) at (axis cs:25,0.22){\scriptsize \textcolor{textgray}{R-OCuLaR}};
       \node (destination) at (axis cs:70,0.19){};
       \draw[->](source)--(destination); 

\end{groupplot}

\end{tikzpicture}
 \vspace{-\baselineskip}
 \caption{
Comparison of OCuLaR with baseline algorithms for the Movielens dataset.  
 \label{fig:Movielens}
 }
 \end{figure}

%%%%%%%%%%%%%%%%%%%%%%%%%%%%%%

We plot the recall@M and MAP@M for the Movielens dataset for varying $M$ in Figure \ref{fig:Movielens}. We see that OCuLaR and R-OCuLaR are consistently better or at least as good as the other recommendation techniques. We summarize the results for the other datasets{\footnote{\scriptsize Netflix dataset is not included because not all baselines can be run for very large datasets}} in Table \ref{tab:results}. 
Across all datasets the OCuLaR variants are either the best or the second-best performing algorithm (together with wALS). OCuLaR has the advantage of providing interpretable recommendations, an aspect that is compromised when using other OCCF approaches such as wALS or BPR. OCuLaR is also significantly better than the user- and item-based algorithms, its interpretable competitors. This is not surprising because the user- and item-based approaches both consider only similarities in either the user or the item space, whereas OCuLaR can discover more complex structures in the joint user-item space. 
%Compared to collaborative filtering techniques, our approach always outperforms the item-to-item approach. It is also significantly better than the user-to-user approach in the Movielens dataset, and it trails only slightly behind it in the client-product dataset.
%One can view the co-clustering approach as a way of meshing together the item-to-user and the user-to-user approaches. 
%More importantly, because our technique uses matrix factorization principles it inherits the robustness and scalability of such approaches. R: not sure if we should say this, MF techniques are not scalable in general, we use a very specific property of the loss function to obtain the scalability we have
%Recall that the traditional collaborative filtering techniques do not scale well under large datasets, because they are plagued by the k-Nearest-Neighbor computation, which can be very costly when dealing with Big Data.
% R: say this later, when discussing scalability?

%%%%%%
\subsection{OCuLaR parameters}
\begin{figure}[!h]
\centering
\begin{tikzpicture}    

 \begin{groupplot}[group style={group size=1 by 6,xlabels at = edge bottom, xticklabels at=edge bottom,vertical sep =0.15cm}, height=2.8cm,width=6.5cm, ylabel style={align=center,rotate=-90, xshift=-1cm},
  legend style={
        %cells={anchor=north},
        at = {(0.5,1.03)}, anchor=south,legend columns=3,
        %legend pos=outer north,
        xlabel=K,
    }
 ]
  \nextgroupplot[ylabel=recall]%@50 items]
  \addplot +[] table[x index=0,y index=1]{./data/movielens_comprop_lam0.dat};
  \addlegendentry{\tiny $\lambda=0$};
  \addplot +[] table[x index=0,y index=1]{./data/movielens_comprop_lam30.dat};
  \addlegendentry{\tiny $\lambda=30$};
  \addplot +[] table[x index=0,y index=1]{./data/movielens_comprop_lam100.dat};
  \addlegendentry{\tiny $\lambda=100$};
    \nextgroupplot[ylabel=users \\ in co-cluster]
    \addplot +[] table[x index=0,y index=6]{./data/movielens_comprop_lam0.dat};
     \addplot +[] table[x index=0,y index=6]{./data/movielens_comprop_lam30.dat};
     \addplot +[] table[x index=0,y index=6]{./data/movielens_comprop_lam100.dat};

    \nextgroupplot[ylabel=items \\ in co-cluster]
    \addplot +[] table[x index=0,y index=7]{./data/movielens_comprop_lam0.dat};
     \addplot +[] table[x index=0,y index=7]{./data/movielens_comprop_lam30.dat};
     \addplot +[] table[x index=0,y index=7]{./data/movielens_comprop_lam100.dat};

      \nextgroupplot[ylabel=co-cluster\\ densities]
    \addplot +[] table[x index=0,y index=3]{./data/movielens_comprop_lam0.dat};
     \addplot +[] table[x index=0,y index=3]{./data/movielens_comprop_lam30.dat};
      \addplot +[] table[x index=0,y index=3]{./data/movielens_comprop_lam100.dat};
    %
    % \nextgroupplot[ylabel=co-clusters an \\ item is in]
    % \addplot +[] table[x index=0,y index=4]{./data/movielens_comprop_lam0.dat};
    %   \addplot +[] table[x index=0,y index=4]{./data/movielens_comprop_lam30.dat};
    %  \addplot +[] table[x index=0,y index=4]{./data/movielens_comprop_lam100.dat};
    
    % % %
    % \nextgroupplot[ylabel=co-clusters a \\ user is in]
    % \addplot +[] table[x index=0,y index=5]{./data/movielens_comprop_lam0.dat};
    %  \addplot +[] table[x index=0,y index=5]{./data/movielens_comprop_lam30.dat};
    %  \addplot +[] table[x index=0,y index=5]{./data/movielens_comprop_lam100.dat};

  \end{groupplot}

\end{tikzpicture}
\vspace{-\baselineskip}
\caption[]{\label{fig:comrecml}
Recall and co-cluster metrics for varying values of the OCuLaR parameters $K$ and $\lambda$.
}
\end{figure}
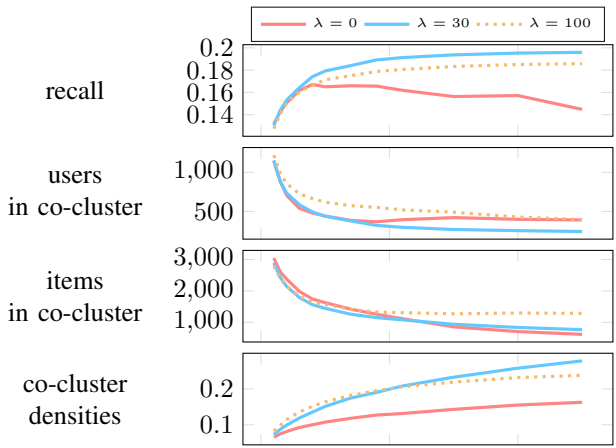

We briefly discuss how to set the parameters for OCuLaR and their impact. Recall that OCuLaR expects the number of co-clusters $K$ and the regularization parameter $\lambda$. Values for $K$ and $\lambda$ are chosen such that the recommendation performance is optimized for a particular metric, as determined via cross-validation. 
%We emphasize that we do not set the parameters such that only the accuracy is maximized, because one of the primary goals of OCuLaR is to ensure high interpretability.
Figure \ref{fig:comrecml} shows the impact of the parameter values on the recommendation performance and on various co-cluster properties on the MovieLens dataset.
The top panel shows the recall@50 items. The graph demonstrates that either too little ($\lambda=0$) or too much regularization ($\lambda=100$) can hurt the recommendation accuracy.  
 $K$ may be selected in such a way to ensure that the size of the co-clusters is neither too big nor too small, and also that each user or item does not belong to too many co-clusters.%; as those are factors that affect the interpretability of the model. 
  The size of the co-clusters can depend on application-specific criteria, such as, for example, that a co-cluster should contain at least 100 users.
For the specific example, a value of $K$ in the range between 100-200 would be adequate to ensure good prediction and avoid excessively big, thus not dense, co-clusters.

%%%%%%%%%%%%%%%%%%%%%%%%%%%%%%
% How to determine parameters for OCuLaR
%%%%%%%%%%%%%%%%%%%%%%%%%%%%%%
%\vspace{-\baselineskip}

%%%%%%%%%%%%%%%%%%%%%%%%%%%%%%%%%%%%%%%%%%%%%%%%%%%%%%%%%%%%%%%%%%%%%%%%%%%%%%%%%

%%%%
\subsection{Scalability}

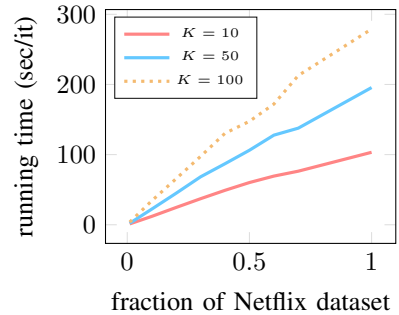
\begin{figure}[!ht]
\centering
\begin{tikzpicture}    

\begin{groupplot}[
width=0.3\textwidth,
group style={group size=1 by 1,
%xticklabels at=edge bottom, xlabels at=edge bottom,
%yticklabels at=edge left, ylabels at=edge left,
horizontal sep=0.8cm,
vertical sep=0.8cm,
},
 every axis y label/.style=
            {at={(ticklabel cs:0.5)},rotate=90,anchor=near ticklabel},
]
\nextgroupplot[legend style={legend pos=north west,},ylabel={running time (sec/it)},xlabel = fraction of Netflix dataset]
\addplot +[] table[x index=0,y index=1]{./data/scaling_netflix.dat};
\addlegendentry{\tiny $K=10$};
\addplot +[] table[x index=0,y index=2]{./data/scaling_netflix.dat};
\addlegendentry{\tiny $K=50$};
\addplot +[] table[x index=0,y index=3]{./data/scaling_netflix.dat};
\addlegendentry{\tiny $K=100$};

%\nextgroupplot[xlabel=number of threads]
%\addplot +[mark=x, draw=none] table[x index=0,y index=1]{./data/scaling_netflix_threads.dat};
\end{groupplot}
\end{tikzpicture}
\vspace{-.5\baselineskip}
\caption{\label{fig:scale} Running time per iteration of the OCuLaR algorithm applied to an increasing fraction of the Netflix dataset. OCuLaR exhibits linear scalability.} %Right: Running time per iteration on the Netflix dataset as a function of the the number of threads.}
\end{figure}

In this section, we investigate the scalability of the OCuLaR algorithm. We consider the full Netflix dataset, consisting of  $100,480,507$ ratings from $480,189$ users on $17,770$ movie titles. As before, we take the ratings $\geq 3$ as positive examples, i.e., $r_{ui}=1$. 
 As analyzed in Section \ref{sec:impdet}, the training time required by the OCuLaR algorithm is essentially linear in the number of positive examples $|\{(u,i)\colon r_{ui}=1\}|$, and linear in the number of co-clusters $K$. In Figure \ref{fig:scale} we plot the running time per iteration for increasing fractions of the Netflix dataset (i.e., non-zero entries), chosen uniformly from the whole Netflix dataset. 
 We see that the training time is indeed linear in the number of positive examples $|\{(u,i)\colon r_{ui}=1\}|$ and linear in the number of co-clusters $K$.

%OCuLaR can also easily be parallelized because for an item (user) update cycle, each item (user) factor can be updated independently. Therefore independent threads can update a fraction of the user and item factors. Figure \ref{fig:scale} depicts the reduction in running time that can be realized by the corresponding parallel OCuLaR implementation as a function of the number of threads. 

\vspace{-\baselineskip}

%%%%%%

\subsection{GPU implementation}

\begin{figure}[!h]
\centering
\includegraphics[width=.75\linewidth]{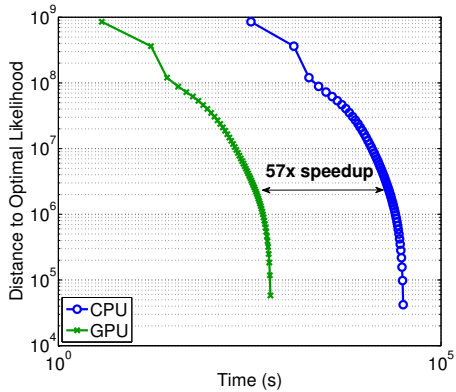}
\vspace{-.5\baselineskip}
\caption{Distance to optimal training likelihood vs time (Netflix dataset, K=200). The GPU implementation is 57 times faster than the CPU.}
\label{fig:gpu_netflix}
\end{figure}

\noindent\textbf{Speedup:}
Here we compare the performance of the CPU implementation with the GPU implementation for the Netflix dataset. The CPU implementation was written in C++ and utilizes the boost library to accelerate certain parts of the computations (such as the evaluation of inner products). It was executed on an Intel Xeon CPU with a clock speed of 2.40GHz. The GPU implementation was written in CUDA C/C++ and executed on a GeForce GTX TITAN X device. In Figure \ref{fig:gpu_netflix} we plot the complete time progression of the training phase, in which OCuLaR minimizes the likelihood of its loss function. The GPU implementation can achieve the same training accuracy over 50 times faster than the CPU implementation. 

\begin{figure}[!h]
\centering
\includegraphics[width=1\linewidth]{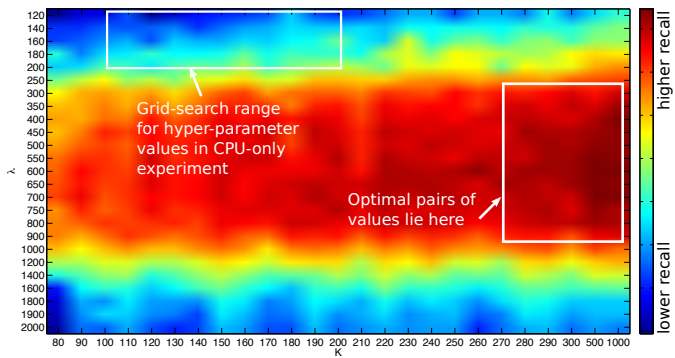}
\vspace{-1.5\baselineskip}
\caption{Grid search for hyper-parameters $(K,\lambda)$ of the algorithm. The heatmap shows the recall@50 products for the \textit{IBM-B2B} dataset. 'Hot' areas resulted in higher recall. One can benefit from the accelarated execution on a GPU to perform a more fine-grained search of the hyper-parameter space.}
\label{fig:gpu_grid}
\end{figure}

\noindent\textbf{Grid search:} The GPU implementation can also help accelerate the grid-search process for learning the algorithm hyper-parameters, namely the regularization parameter $\lambda$ and the number of co-clusters $K$.
This process can be further accelerated by scaling out across a GPU-enabled cluster. In Figure \ref{fig:gpu_grid} we show the result of a very fine grid-search over 625 different parameter pairs for the \textit{IBM-B2B} dataset. The parameter pairs were distributed using Apache Spark across a cluster of 8 machines, each fitted with a NVIDIA Quadro M4000 GPU. Each Spark worker performed training for the parameters assigned to it by calling compiled CUDA code via the Python-C API. The entire grid-search using 8 GPUs took around 8 minutes. On a single CPU this would have required more than 2 days. In Figure \ref{fig:gpu_grid}, the two rectangles show the range of `optimal' hyper-parameters lies, and the grid-search range used in the CPU-only experiment (Fig. \ref{fig:Movielens}). Therefore, the recall reported in the previous experiments could have been improved even further had one used a more exhaustive hyper-parameter grid-search method. This shows the benefit of having a very fast GPU implementation. 

\vspace{-.5\baselineskip}

%%%%%%%%%%%%%%%%%%%%%%%%%%%%%%%%%%%%%%%%%%%%%%%%%%%%%%%%%%%%%%%%%%%%%%%%%%%%%%%%%

\section{Deployment and User Feedback}
\label{sec:inddeploy}

We used our algorithm in a B2B recommender system of our institution. These recommendations are not offered directly to the clients of our  institution, but rather to our sales teams. The salesperson responsible for an account examines the  recommendations provided and decides whether to act on the recommendation and approach the client. This decision is based on the reasoning provided and on the salesperson's own experience through past interaction with the client.

\begin{figure}[!h]
\centering
\includegraphics[width=1.05\linewidth]{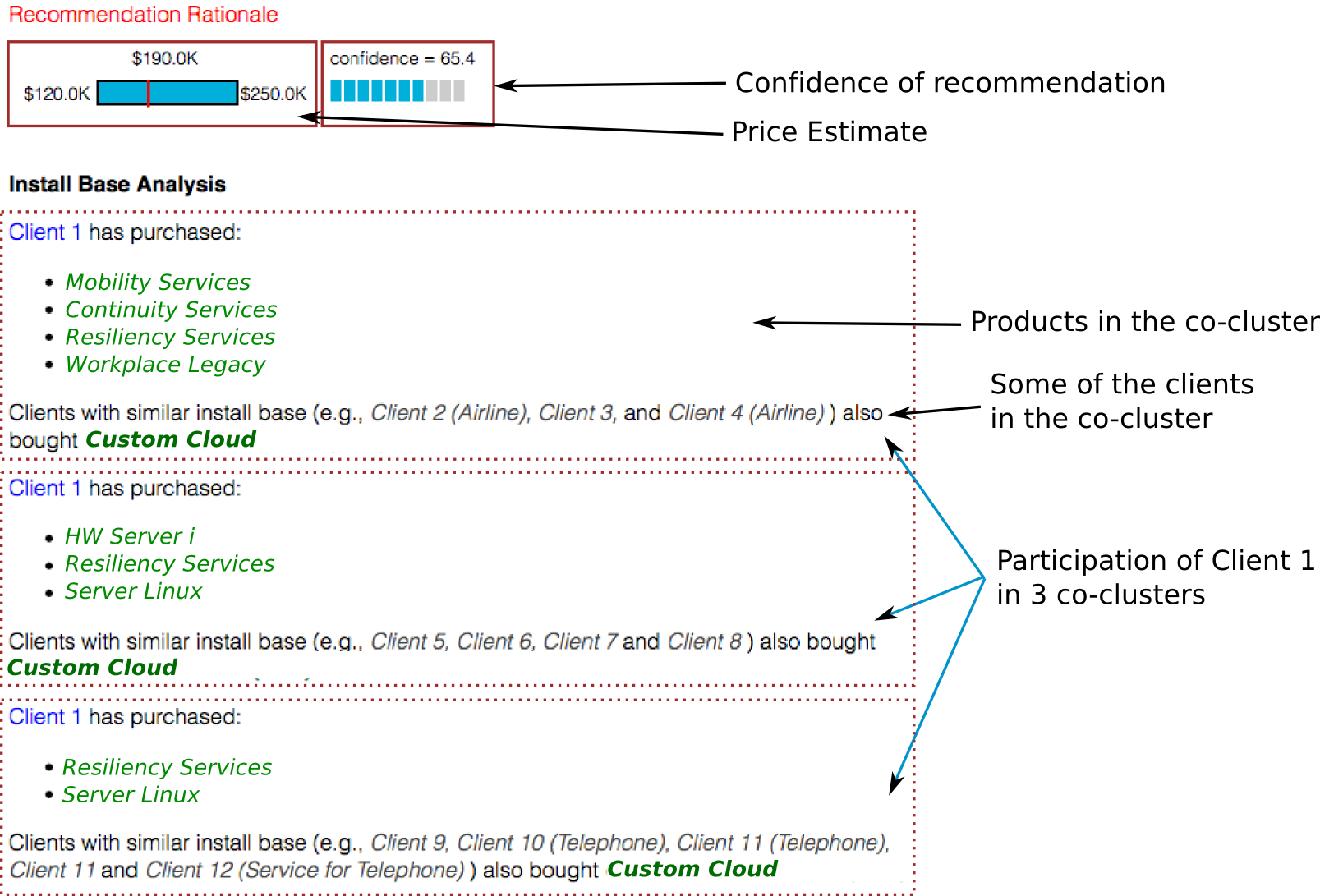}
\caption{
Example from industrial deployment (client names are suppressed). Product ``Custom Cloud'' is recommended to Client 1  because the client is affiliated with three co-clusters. 
Co-cluster 1 shows an affinity of Client 1 with several airlines, and in Co-cluster 3 with telco companies. 
}
\label{fig:example_s}
\end{figure}

The textual output of the recommendation, which conveys the corresponding rationale is shown in Figure \ref{fig:example_s}. 
For reasons of anonymity, the names of the companies/clients belonging to each co-cluster have been omitted. 

%\vspace{-.7\baselineskip}

In the example, we see that the service ``Custom Cloud" is recommended for ``Client 1" with a confidence of $65.4\%$. The reasoning explains that Client 1 belongs to three co-clusters that have bought the same service, as well as products similar to the ones that Client 1 has already purchased. Based on the co-clusters discovered, the  interface also presents a price estimate of the potential business deal, based on historical purchases of the same product by the related clients belonging to the co-clusters discovered.

During the deployment of the OCuLaR algorithm in our organization, we interacted with many sellers that used the platform. Sellers expressed satisfaction about the reasoned aspect of the recommendations. An interesting comment that we received is that the tool could also constitute an \textit{educational} platform \cite{Cleger2014} for young sellers, because the detailed reasoning can teach them the currently discovered buying patterns. %In an extended version, we plan to provide additional metrics, such as the conversion ratio and the monetary benefits of the industrial deployment. 

\section{Conclusion}
A large body of work on recommender systems and machine learning has focused primarily on the accuracy of prediction rather than on interpretability. This work explicitly addresses the aspect of interpretability by enabling the detection of overlapping co-clusters that can be easily visualized and transcribed into a textual description. 
The methodology presented is interpretable, scalable, and does not sacrifice accuracy of prediction.
We demonstrated also an efficient GPU implementation which allows the process to be further accelerated by more than 50 times and helps to better explore the hyper-parameter space of the algorithm.

Finally, even though the focus of this work was on recommender systems, we feel that the algorithm presented can be used for solving large co-clustering problems in other disciplines as well, including community discovery in social networks \cite{palla2005uncovering}, or for the analysis of gene expression data \cite{prelic2006systematic}. 

\medskip
\noindent
% \scriptsize{
\textbf{Acknowledgements:}
The research leading to these results has received funding from the European Research Council under the European Union's Seventh Framework Programme (FP7/2007-2013) / ERC grant agreement no. 259569.
% }
%\scriptsize{
\bibliographystyle{abbrv}
\bibliography{database}
%}

\end{document}